\documentclass[showkeys,superscriptaddress,floatfix,prd,10pt,aps]{revtex4-2}
\usepackage{graphicx,epstopdf}
\usepackage[caption=false]{subfig}
\usepackage{appendix}
\usepackage[T1]{fontenc}
\usepackage{lmodern}
\usepackage[dvipsnames,x11names]{xcolor}
\usepackage[colorlinks=true,linkcolor=NavyBlue,citecolor=ForestGreen,urlcolor=NavyBlue]{hyperref}
\usepackage[sort&compress]{natbib}
\usepackage{lipsum}
\usepackage{morefloats}
\usepackage[pdf]{pstricks}
\usepackage{amsmath}
\usepackage{amssymb}
\usepackage{amsfonts}
\usepackage{rotating}
\usepackage{cancel}
\usepackage{mathtools}
\usepackage{bbm}
\usepackage{dsfont}
\usepackage{bbold}
\usepackage{multirow}
\usepackage{ulem}
\usepackage{physics}
\usepackage{orcidlink}
\usepackage{colortbl}

\def\pslash{p\!\!\!\slash }
\def\p_1slash{p1\!\!\!\slash }
\def\p_2slash{p2\!\!\!\slash }
\def\qslash{q\!\!\!\slash }
\def\xslash{x\!\!\!\slash }
\def\eslash{\varepsilon\!\!\!\slash }

\begin{document}

\title{Electromagnetic tomography of spin-$\frac{3}{2}$ hidden-charm strange pentaquarks}

\author{Ula\c{s}~\"{O}zdem\orcidlink{0000-0002-1907-2894}}%
\email[]{ulasozdem@aydin.edu.tr }
\affiliation{ Health Services Vocational School of Higher Education, Istanbul Aydin University, Sefakoy-Kucukcekmece, 34295 Istanbul, T\"{u}rkiye}

\begin{abstract}
Understanding how quarks are spatially arranged inside exotic pentaquarks remains one of the key open problems in contemporary hadron spectroscopy. The electromagnetic multipole moments of hadrons  provide a direct probe of their internal quark--gluon geometry and spatial charge distributions. Motivated by this, we employ QCD light-cone sum rules to compute the magnetic dipole, electric quadrupole, and magnetic octupole moments of the $J^P = 3/2^-$ pentaquark with strangeness $S = -1$. Five distinct diquark--diquark--antiquark interpolating currents are constructed to explore possible internal configurations. The resulting electromagnetic moments exhibit pronounced sensitivity to the underlying quark arrangement: magnetic dipole moments range from $-2.28\mu_N$ to $+3.36\mu_N$, establishing this observable as a key discriminator among configurations with identical quantum numbers. Nonzero electric quadrupole and magnetic octupole moments indicate clear deviations from spherical symmetry, while a detailed decomposition shows that light quarks dominate the magnetic response and the charm quark drives quadrupole deformation. These findings position electromagnetic multipole moments as quantitative and discriminating probes of exotic hadron structure, providing concrete benchmarks for forthcoming LHCb, Belle II, and lattice QCD studies.
\end{abstract}

%\keywords{Electromagnetic form factors, diquark-diquark-antiquark configuration, hidden-charm strange pentaquarks, QCD light-cone sum rules}
%\date{\today}
\maketitle

\section{Motivation}\label{motivation}

The concept of hadronic states beyond the conventional quark-antiquark mesons and three-quark baryons has been proposed for many decades. Experimental verification of such exotic configurations was first achieved in 2003 with the discovery of the $X(3872)$ by the Belle Collaboration~\cite{Belle:2003nnu}. Due to its properties that cannot be explained within the framework of standard mesons, this state is widely regarded as the first confirmed tetraquark candidate. Since then, numerous exotic hadrons featuring nontraditional quark arrangements have been observed, and the variety of these states has steadily expanded. These findings have stimulated extensive theoretical and experimental studies aimed at understanding the nature and internal structure of these unconventional multi-quark systems, which in turn provide valuable information on the dynamics of strong interactions at low energies. A range of theoretical interpretations has been proposed, including compact multiquark models, hadronic molecules, and explanations based on kinematic effects, among others~\cite{Esposito:2014rxa, Esposito:2016noz, Olsen:2017bmm, Lebed:2016hpi, Nielsen:2009uh, Brambilla:2019esw, Agaev:2020zad, Chen:2016qju, Ali:2017jda, Guo:2017jvc, Liu:2019zoy, Yang:2020atz, Dong:2021juy, Dong:2021bvy, Chen:2022asf, Meng:2022ozq, Alasiri:2025roh, Brambilla:2025xma}. Together, these efforts have significantly advanced our understanding of exotic hadronic matter and continue to guide ongoing research in hadron spectroscopy.

In 2015, the LHCb Collaboration achieved a significant milestone in the study of exotic multi-quark hadrons by reporting two pentaquark states, denoted $P_{\psi}^{N}(4380)$ and $P_{\psi}^{N}(4450)$, observed in the $J/\psi p$ invariant mass spectrum from the decay $\Lambda^0_b \to J/\psi p K^-$. Later, with an updated analysis incorporating data from both Run I and Run II, the LHCb Collaboration refined their observations of the $J/\psi p$ spectrum in $\Lambda_b \to J/\psi p K$ decays. This analysis revealed a new pentaquark, $P_{\psi}^{N}(4312)$, and resolved the previously reported $P_{\psi}^{N}(4450)$ into two distinct states, labeled $P_{\psi}^{N}(4440)$ and $P_{\psi}^{N}(4457)$. The status of $P_{\psi}^{N}(4380)$, however, remains inconclusive in the updated study, leaving its existence uncertain. Since these states appear in the $J/\psi p$ channel, their minimal quark content is expected to be $uudc\bar c$.   Building on these discoveries, the LHCb Collaboration extended the search for hidden-charm pentaquarks with strangeness. In 2020, they reported evidence for a candidate state, $P^{\Lambda}_{\psi s}(4459)$, in the $J/\psi \Lambda$ invariant mass spectrum from the decay $\Xi^-_b \to J/\psi \Lambda K^-$. Subsequently, in 2022, another candidate, $P^{\Lambda}_{\psi s}(4338)$, was observed in the same channel. The quark composition of these $P^{\Lambda}_{\psi s}$ states is generally assumed to be $udsc\bar c$, consistent with their decay products. The masses and decay widths of these pentaquark states can be seen in Refs.~\cite{LHCb:2015yax, LHCb:2019kea, LHCb:2020jpq, LHCb:2022ogu}. Very recently, the Belle Collaboration reported evidence for a $P^{\Lambda}_{\psi s}$ state, achieving a significance of 3.3$\sigma$ when both statistical and systematic uncertainties are taken into account. The measured mass and width of this state are $(4471.7 \pm 4.8 \pm 0.6)$ MeV and $(21.9 \pm 13.1 \pm 2.7)$ MeV, respectively~\cite{Belle:2025pey}. Nevertheless, given the limited available data, one cannot exclude the possibility that the observed signal originates from two separate states. The study further indicates that the observed state could correspond to a different state in the same mass region \cite{Clymton:2025hez}. This possibility points to overlapping or closely spaced resonances, emphasizing the need for careful experimental and theoretical analysis to accurately determine the nature and quantum numbers of the observed state.

The spin and parity of the hidden-charm pentaquark states have not yet been experimentally established. Despite significant theoretical and experimental efforts following their initial discovery, the underlying nature and internal structure of these exotic states remain largely unresolved. This lack of definitive information underscores the importance of further detailed studies aimed at uncovering their properties. A thorough understanding of these systems requires the investigation of various observables that probe the internal dynamics of the states. Quantities such as decay constants, branching ratios, and transition form factors provide critical insights into their quark-gluon composition and interaction patterns. Among these, electromagnetic properties, and in particular magnetic moments, constitute especially valuable probes. Magnetic moments not only quantify the distribution of charge and magnetization within the hadron but also offer a direct window into the spatial arrangement and spin alignment of constituent quarks. By analyzing these moments, one can infer key aspects of the internal configuration, such as the role of different quark flavors and possible deformation patterns, thereby complementing other experimental and theoretical investigations and guiding future studies aimed at resolving the true nature of hidden-charm pentaquarks. 

\begin{widetext}
 
 \begin{figure}[htp]
\centering
\includegraphics[width=1.0\textwidth]{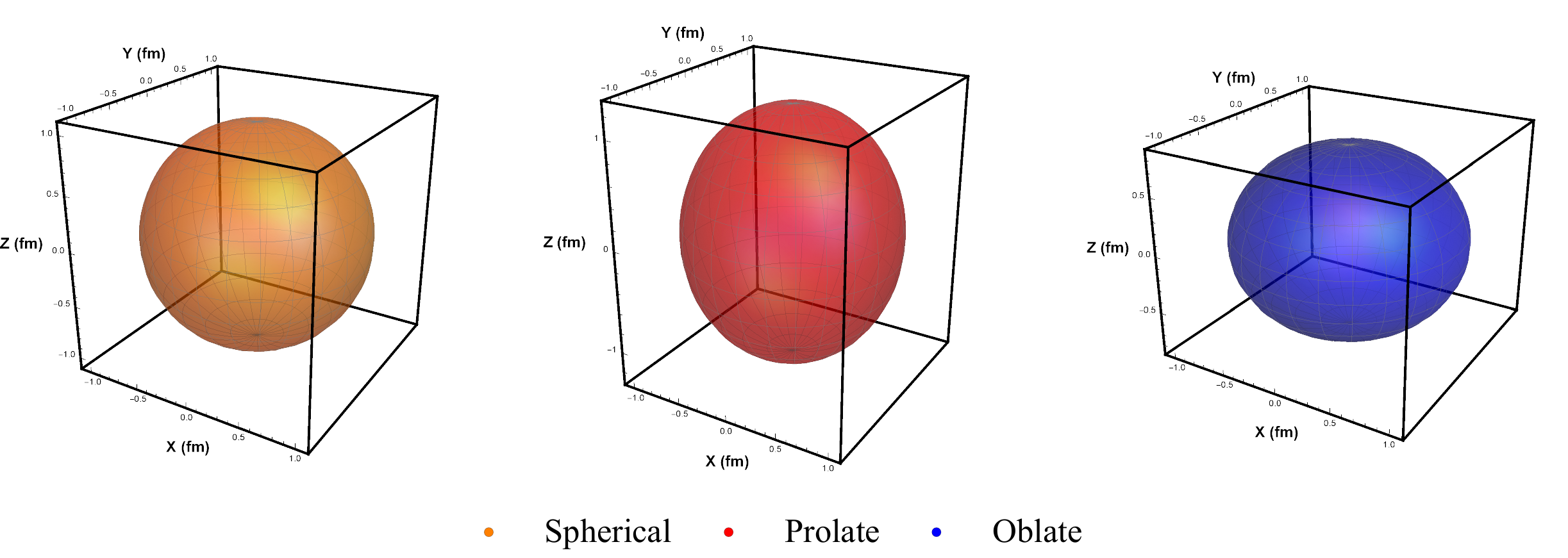} 
\caption{Illustrates the charge distributions corresponding to intrinsic quadrupole moments ($\mathcal{Q}$): spherical ($\mathcal{Q}=0$), positive (prolate, $\mathcal{Q}>0$), and negative (oblate, $\mathcal{Q}<0$). Similarly, a structural asymmetry is observed for the octupole moment ($\mathcal{O}$).}
    \label{Quadrupole_Shapes}
  \end{figure}

\end{widetext}

Inspired by these developments, the growing interest in hidden-charm pentaquark states, and the open questions that still need to be addressed, in this work, we present a systematic analysis of the magnetic dipole, electric quadrupole, and magnetic octupole moments of hidden-charm pentaquarks with strangeness $S= -1$, employing the framework of QCD light-cone sum rules~\cite{Chernyak:1990ag, Braun:1988qv, Balitsky:1989ry}. Our study considers five distinct interpolating currents constructed in diquark-diquark-antiquark configurations, chosen for their expected effectiveness in coupling to pentaquark states with $J^P = 3/2^-$. Electromagnetic interactions provide a fundamental and non-invasive probe of the spatial and dynamical properties of quarks and gluons inside hadrons. These interactions are described through elastic and inelastic multipole form factors, which encode detailed information on the internal geometry of hadrons, including their shape, size, and charge and current distributions (See Fig. \ref{Quadrupole_Shapes}). The multipole moments, as leading-order responses to external electromagnetic fields, offer particularly direct insight into the structural characteristics of hadronic states.    
The topic has attracted growing attention in the literature due to the valuable insights it provides, with numerous studies employing a range of models and techniques to investigate the electromagnetic multipole moments of hidden-charm and hidden-bottom pentaquarks~\cite{Wang:2016dzu, Ozdem:2018qeh, Ortiz-Pacheco:2018ccl, Xu:2020flp, Ozdem:2021btf, Ozdem:2021ugy, Li:2021ryu, Ozdem:2023htj, Wang:2023iox, Ozdem:2022kei, Gao:2021hmv, Ozdem:2024rch, Guo:2023fih, Ozdem:2022iqk, Wang:2022nqs, Wang:2022tib, Ozdem:2024jty, Li:2024wxr, Li:2024jlq, Ozdem:2024yel, Ozdem:2024rqx, Mutuk:2024ltc, Mutuk:2024jxf, Ozdem:2024usw, Ozdem:2025fks, Zhu:2025abk, Ozdem:2025ncd}.   

The paper is structured as follows. In the next section, we describe the methodology used to construct suitable interpolating currents for the $P^{\Lambda}_{\psi s}$ pentaquark and derive the corresponding sum rules for the physical observables of interest. The numerical analysis and associated discussions are presented in Sec.~\ref{numerical}. A detailed structural interpretation of the results and the corresponding experimental signatures are presented in Sec.~\ref{sec:discussion}. Finally, we conclude with a summary of the main findings and implications of this work in Sec.~\ref{summary}.

\begin{widetext}
 
\section{QCD light-cone sum rules setup}\label{formalism}

The first step in the calculation involves the definition of an appropriate correlation function that encapsulates the relevant quantum numbers and interpolating currents of the system under study. For the analysis of the electromagnetic multipole moments of the $P^{\Lambda}_{\psi s}$ state, the correlation function is constructed as follows:
 
\begin{eqnarray} \label{edmn01}
T_{\mu \nu}(p,q)&=&i\int d^4x e^{ip \cdot x} \langle0|T\left\{J_{\mu}(x)\bar{J}_{\nu}(0)\right\}|0\rangle _F \, .
\end{eqnarray}
Here, $F$ represents the external electromagnetic field that induces the interaction responsible for the multipole transitions, while $J_{\mu}(x)$ denotes the interpolating current corresponding to the $P^{\Lambda}_{\psi s}$ state. The possible interpolating currents that couple to the $P^{\Lambda}_{\psi s}$ state are given as:
\begin{eqnarray}
J_{\mu}^{1}(x)&=&\frac{\mathcal A }{\sqrt{2}} \bigg\{ \big[ {u}^T_d(x) C \gamma_5 {s}_e(x) \big] \big[ {d}^T_f(x) C \gamma_\mu c_g(x)\big]  -
\big[ {d}^T_d(x) C \gamma_5 {s}_e(x) \big] \big[ {u}^T_f(x) C \gamma_\mu c_g(x)\big]  
\bigg\}  C  \bar{c}^{T}_{c}(x) \, , \\
%%%%%%%%%%%%%%%%%%%%%%%%%%%%%%%%%%%%%%%%%%%%%%%%%%
J_{\mu}^{2}(x)&=& \frac{\mathcal A }{\sqrt{2}} \bigg\{\big[ {u}^T_d(x) C \gamma_\mu {s}_e(x) \big] \big[ {d}^T_f(x) C \gamma_5 c_g(x)\big]  -
\big[ {d}^T_d(x) C \gamma_\mu {s}_e(x) \big] \big[ {u}^T_f(x) C \gamma_5 c_g(x)\big]  \bigg\} C  \bar{c}^{T}_{c}(x)  \, , \\
%%%%%%%%%%%%%%%%%%%%%%%%%%%%%%%%%%%%%%%%%%%%%%%%%%%
J_{\mu}^{3}(x)&=& \frac{\mathcal A }{\sqrt{2}} \bigg\{ \big[ {u}^T_d(x) C \gamma_\mu {s}_e(x) \big] \big[ {d}^T_f(x) C \gamma_\alpha c_g(x)\big]  -
\big[ {d}^T_d(x) C \gamma_\mu {s}_e(x) \big] \big[ {u}^T_f(x) C \gamma_\alpha c_g(x)\big]  \bigg\} \gamma_5 \gamma^\alpha C  \bar{c}^{T}_{c}(x) \, , \\
%%%%%%%%%%%%%%%%%%%%%%%%%%%%%%%%%%%%%%%%%%%%%%%%%%%
J_{\mu}^{4}(x)&=& \frac{\mathcal A }{\sqrt{2}} \bigg\{ \big[ {u}^T_d(x) C \gamma_\alpha {s}_e(x) \big] \big[ {d}^T_f(x) C \gamma_\mu c_g(x)\big]  -
\big[ {d}^T_d(x) C \gamma_\alpha {s}_e(x) \big] \big[ {u}^T_f(x) C \gamma_\mu c_g(x)\big]  \bigg\} \gamma_5 \gamma^\alpha C  \bar{c}^{T}_{c}(x) \, , \\
%%%%%%%%%%%%%%%%%%%%%%%%%%%%%%%%%%%%%%%%%%%%%%%%%%%
J_{\mu}^{5}(x)&=&\mathcal A  \bigg\{ \big[ {u}^T_d(x) C \gamma_5 {d}_e(x) \big] \big[ {s}^T_f(x) C \gamma_\mu c_g(x)\big] \bigg\}  C  \bar{c}^{T}_{c}(x) \, ,
 \end{eqnarray} 
where $\mathcal{A} = \varepsilon_{abc}\varepsilon_{ade}\varepsilon_{bfg}$, with $a$, $b$, $c$, $d$, $e$, $f$, and $g$ denoting color indices, and $C$ representing the charge conjugation operator. 
 
The five interpolating currents $J^{1}_{\mu}(x)$--$J^{5}_{\mu}(x)$ are constructed to probe distinct configurations within the complex wavefunction of a $J^{P} = \tfrac{3}{2}^{-}$ hidden-charm strange pentaquark. All currents are carefully built to maintain the correct quantum numbers while emphasizing different diquark correlations and spatial arrangements, thereby allowing us to assess the structural sensitivity of electromagnetic observables across a broad spectrum of possible internal organizations. 
A more detailed justification for this quintuple current approach is as follows: $J^{1}_{\mu}(x)$, the scalar--axial-vector current, combines a scalar light-strange diquark $[u^T C \gamma_5 s]$ with an axial-vector charm-light diquark $[d^T C \gamma_\mu c]$. While this configuration primarily emphasizes compact diquark correlations, its specific structure may also allow coupling to molecular-type configurations where distinct meson-baryon components could overlap with this diquark arrangement. 
The current $J^{2}_{\mu}(x)$, the axial-vector--scalar current, reverses the diquark roles compared to $J^{1}_{\mu}$, employing an axial-vector light-strange diquark $[u^T C \gamma_\mu s]$ with a scalar charm-light diquark $[d^T C \gamma_5 c]$. This swapped arrangement probes how the assignment of scalar versus axial-vector character to heavy and light diquarks affects the electromagnetic properties, while maintaining the possibility of overlap with both compact and molecular configurations through different diquark clustering patterns. 
Meanwhile, $J^{3}_{\mu}(x)$ and $J^{4}_{\mu}(x)$ both represent axial-vector--axial-vector configurations but with crucial differences in Lorentz index assignments. $J^{3}_{\mu}$ places the Lorentz index $\alpha$ in the charm-containing diquark $[d^T C \gamma_\alpha c]$, while $J^{4}_{\mu}$ positions it in the light-strange diquark $[u^T C \gamma_\alpha s]$. Both employ the $\gamma_5 \gamma^\alpha$ structure for correct angular momentum coupling. These currents are primarily optimized for probing compact multiquark states with strong color correlations, as their symmetric axial-vector diquark structures favor tightly bound configurations. While the possibility of coupling to specific molecular arrangements cannot be entirely excluded—particularly those with correlated spin structures that might overlap with these diquark combinations—their design makes them most sensitive to genuine compact pentaquark components. 
Finally, $J^{5}_{\mu}(x)$, the unique non-strange--strange separated current, features a scalar non-strange diquark $[u^T C \gamma_5 d]$ coupled to an axial-vector strange-charm diquark $[s^T C \gamma_\mu c]$. This clear separation between non-strange and strange-heavy sectors suggests strong overlap with compact configurations featuring distinct quark sector clustering, while the specific diquark combinations may also permit coupling to certain molecular-type wavefunctions. 
By employing this set of five currents, we effectively scan a comprehensive range of possible internal structures within the hidden-charm strange pentaquark spectrum. The significant variation in predicted multipole moments across these currents provides a quantitative measure of how sensitively these fundamental observables depend on the underlying quark-gluon architecture, revealing the rich structural diversity accessible even within the same quantum number sector, without presupposing the dominance of any particular structural model.   It should be noted that QCD sum rules suggest the scalar and axial-vector diquark configurations to be the most favorable for studying hadronic properties~\cite {Wang:2010sh, Kleiv:2013dta}. Accordingly, the interpolating currents introduced above are constructed in various combinations based on these configurations.

%\subsection{Hadronic framework of the correlation function}

In the hadronic framework, the principle of quark-hadron duality is employed to transform the correlation function into its hadronic representation. This is achieved by inserting a complete set of intermediate states possessing the same quantum numbers as the interpolating currents. The contribution from the ground state, denoted as $P^{\Lambda}_{\psi s}$, is isolated from those of higher-energy states and the continuum, yielding the following expression:

\begin{eqnarray}\label{edmn02}
T^{Had}_{\mu\nu}(p,q)&=&\frac{\langle0\mid J_{\mu}(x)\mid
P^{\Lambda}_{\psi s}(p_2,s)\rangle}{[p_2^{2}-m_{P^{\Lambda}_{\psi s}}^{2}]}\langle P^{\Lambda}_{\psi s}(p_2,s)\mid
P^{\Lambda}_{\psi s}(p_1,s)\rangle_F\frac{\langle P^{\Lambda}_{\psi s}(p_1,s)\mid
\bar{J}_{\nu}(0)\mid 0\rangle}{[p_1^{2}-m_{P^{\Lambda}_{\psi s}}^{2}]} \nonumber\\
&&+ \mbox{ higher states and continuum}.
\end{eqnarray}
Here, $p_1 = p + q$, $p_2 = p$, and $q$ denotes the momentum of the photon. 
The evaluation of this correlation function requires the explicit forms of the matrix elements involving the $P^{\Lambda}_{\psi s}$ state.

The relevant matrix elements are defined as follows:

\begin{eqnarray}
\label{lambdabey}
\langle0\mid J_{\mu}(x)\mid P^{\Lambda}_{\psi s}(p_2,s)\rangle &=&\lambda_{P^{\Lambda}_{\psi s}}u_{\mu}(p_2,s),\\
%\nonumber\\
%%%%%%%%%%%%%%%%%%%%%%%%%%%%%%%%%%%%%%%%%%%%%%%%%%%
\langle {P^{\Lambda}_{\psi s}}(p_1,s)\mid
\bar{J}_{\nu}(0)\mid 0\rangle &=& \lambda_{{P^{\Lambda}_{\psi s}}}\bar u_{\nu}(p_1,s),%\\
\end{eqnarray}
and the electromagnetic transition vertex is given by~\cite{Weber:1978dh,Nozawa:1990gt,Pascalutsa:2006up,Ramalho:2009vc}:

\begin{eqnarray}
%\nonumber\\
%%%%%%%%%%%%%%%%%%%%%%%%%%%%%%%%%%%%%%%%
\langle P^{\Lambda}_{\psi s}(p_2,s)\mid P^{\Lambda}_{\psi s}(p_1,s)\rangle_F &=&-e \,\bar
u_{\mu}(p_2,s)\Gamma_{\mu\nu} u_{\nu}(p_1,s),\label{matelpar}
\end{eqnarray}
with
\begin{align}\label{matelpar1}
 \Gamma_{\mu\nu} &= F_{1}(q^2)g_{\mu\nu}\eslash-
\frac{1}{2m_{P^{\Lambda}_{\psi s}}}
\bigg[F_{2}(q^2)g_{\mu\nu} \eslash\qslash+F_{4}(q^2)\frac{q_{\mu}q_{\nu} \eslash\qslash}{(2m_{P^{\Lambda}_{\psi s}})^2}\bigg]
+
\frac{F_{3}(q^2)}{(2m_{P^{\Lambda}_{\psi s}})^2}q_{\mu}q_{\nu}\eslash ,
\end{align}
where $\lambda_{P^{\Lambda}_{\psi s}}$ denotes the pole residue, $u_{\mu}$ is the Rarita-Schwinger spinor, $\varepsilon$ is the photon polarization vector, and $F_i(q^2)$ are the electromagnetic form factors.

Substituting Eqs.~(\ref{edmn02})--(\ref{matelpar1})  into the correlation function and performing the necessary algebraic manipulations yields the final hadronic representation:

\begin{eqnarray}
\label{final phenpart}
T^{Had}_{\mu\nu}(p,q)&=&\frac{\lambda_{_{P^{\Lambda}_{\psi s}}}^{2}}{[(p+q)^{2}-m_{_{P^{\Lambda}_{\psi s}}}^{2}][p^{2}-m_{_{P^{\Lambda}_{\psi s}}}^{2}]}
\bigg[ 2 \varepsilon_\mu q_\nu \qslash \,F_{1}(q^2) 
+ 2 q_\mu q_\nu \qslash \,F_{2}(q^2)+
\frac{F_{3}(q^2)}{2m_{P^{\Lambda}_{\psi s}}}q_{\mu}q_{\nu}\eslash\qslash\, 
+ \frac{F_{4}(q^2)}{4m_{P^{\Lambda}_{\psi s}}^3}(\varepsilon.p)q_{\mu}q_{\nu}\pslash\qslash \,
\nonumber\\
&+&
\mathrm{other~independent~structures} \bigg].
\end{eqnarray}

For clearer physical interpretation, the form factors $F_i(q^2)$ are related to the magnetic dipole ($G_M$), electric quadrupole ($G_Q$), and magnetic octupole ($G_O$) form factors~\cite{Weber:1978dh,Nozawa:1990gt,Pascalutsa:2006up,Ramalho:2009vc}: 

\begin{eqnarray}
G_{M}(q^2) &=& \left[ F_1(q^2) + F_2(q^2)\right] ( 1+ \frac{4}{5}
\lambda ) -\frac{2}{5} \left[ F_3(q^2)  +
F_4(q^2)\right] \lambda \left( 1 + \lambda \right), \nonumber\\
G_{Q}(q^2) &=& \left[ F_1(q^2) -\lambda F_2(q^2) \right]  -
\frac{1}{2}\left[ F_3(q^2) -\lambda F_4(q^2)
\right] \left( 1+ \lambda \right),  \nonumber \\
 G_{O}(q^2) &=&
\left[ F_1(q^2) + F_2(q^2)\right] -\frac{1}{2} \left[ F_3(q^2)  +
F_4(q^2)\right] \left( 1 + \lambda \right),
\end{eqnarray}
  where $\lambda
= -\frac{q^2}{4m^2_{P^{\Lambda}_{\psi s}}}$. At $q^2=0$, these relations simplify to:
\begin{eqnarray}\label{mqo1}
G_{M}(0)&=&F_{1}(0)+F_{2}(0),\nonumber\\
G_{Q}(0)&=&F_{1}(0)-\frac{1}{2}F_{3}(0),\nonumber\\
G_{O}(0)&=&F_{1}(0)+F_{2}(0)-\frac{1}{2}[F_{3}(0)+F_{4}(0)].
\end{eqnarray}
 
Since the present study focuses on the magnetic dipole, electric quadrupole, and magnetic octupole moments, these electromagnetic multipole moments are expressed in terms of the form factors introduced above. The magnetic dipole ($\mu_{P^{\Lambda}_{\psi s}}$), electric quadrupole ($Q_{P^{\Lambda}_{\psi s}}$), and magnetic octupole ($O_{P^{\Lambda}_{\psi s}}$) moments can then be obtained from the corresponding form factors through the following procedure:
 
 \begin{eqnarray}
\mu_{P^{\Lambda}_{\psi s}}&=G_{M}(0)\,\frac{e}{2m_{P^{\Lambda}_{\psi s}}},\\
Q_{P^{\Lambda}_{\psi s}}&=G_{Q}(0) \,\frac{e}{m_{P^{\Lambda}_{\psi s}}^2}, \\
O_{P^{\Lambda}_{\psi s}}&=G_{O}(0)\,\frac{e}{2m_{P^{\Lambda}_{\psi s}}^3}.
\label{mqo2}
\end{eqnarray}

From Ref.~\cite{Wang:2025fqh}, it is observed that the masses of the $P^{\Lambda}_{\psi s}$ pentaquark states extracted from five different interpolating currents fall within a narrow region. In general, an interpolating current may couple to one or several hadronic states sharing the same quantum numbers, since such couplings are not prohibited by symmetries. Conversely, a hadron can contain multiple Fock components, each potentially coupling to a current with identical quantum numbers. 
These findings imply that a pentaquark state may accommodate different internal configurations, or that distinct pentaquark structures with similar masses coexist within the same mass region. If the mass splittings among these states are sizable, the expressions in Eqs.~(\ref{edmn02})--(\ref{mqo2}) apply to the lowest-mass state, and the magnetic dipole  moments obtained from different interpolating currents are expected to converge. In contrast, if the masses are nearly degenerate, Eq.~(\ref{edmn02}) yields a residue-squared--weighted average of the magnetic moments of the overlapping states. In the latter case, substantial differences between the intrinsic magnetic dipole moments of these nearly degenerate states will manifest as variations among the results obtained from different interpolating currents, reflecting their distinct coupling strengths to the physical configurations.

%\subsection{QCD framework of the correlation function}

The QCD representation of the correlation function is derived by evaluating it in terms of quark and gluon degrees of freedom. This involves applying the operator product expansion (OPE) near the light-cone, which systematically incorporates both the perturbative and non-perturbative dynamics of the photon through its distribution amplitudes (DAs).

Applying Wick's theorem to contract the quark fields, the correlation function is expressed in terms of the relevant quark propagators. For the specific case of the $J_\mu^1$ interpolating current, we obtain the following explicit form:

\begin{eqnarray}
\label{QCD1}
T^{QCD}_{\mu\nu}(p,q)&=&-i \, \mathcal A \, \mathcal{A^\prime}\int d^4x e^{ip\cdot x}\langle 0|
\Bigg\{
  \mbox{Tr}\Big[  \gamma_5 S_s^{ee^\prime}(x) \gamma_5  \widetilde S_u^{dd^\prime \mathrm{T}}(x) \Big]
 \mbox{Tr}\Big[ \gamma_\mu S_c^{gg^\prime}(x) \gamma_\nu \widetilde  S_d^{ff^\prime \mathrm{T}}(x) \Big] 
 %%%%%%%%%%%%%%%%%%%%%%%%%%%%%%%%%%%%%%%%%%%%%%%%%%
 \nonumber\\
&&
+ \mbox{Tr}\Big[  \gamma_5 S_s^{ee^\prime}(x) \gamma_5  \widetilde S_d^{dd^\prime \mathrm{T}}(x) \Big]
 \mbox{Tr}\Big[ \gamma_\mu S_c^{gg^\prime}(x) \gamma_\nu \widetilde S_u^{ff^\prime \mathrm{T}}(x) \Big]  
 %%%%%%%%%%%%%%%%%%%%%%%%%%%%%%%%%%%%%%%%%%%%%%%%%%
\Bigg \} \Big( \widetilde S_c^{c^{\prime}c \mathrm{T}} (-x)  \Big)
|0 \rangle_F ,
\end{eqnarray}
where $\mathcal A \mathcal{A^\prime}= \frac{1}{2} \, \varepsilon^{abc}\varepsilon^{a^{\prime}b^{\prime}c^{\prime}}\varepsilon^{ade}
\varepsilon^{a^{\prime}d^{\prime}e^{\prime}}\varepsilon^{bfg}
\varepsilon^{b^{\prime}f^{\prime}g^{\prime}}$; and 
$\widetilde{S}_{c(q)}^{ij}(x)=CS_{c(q)}^{ij\mathrm{T}}(x)C,
$ 
with $C$ and $T$ being the charge conjugation and transpose of the
operator, respectively.

In Eq.~(\ref{QCD1}), $S_{q}(x)$ and $S_{c}(x)$ represent the light and charm quark propagators. Their explicit expressions are given by~\cite{Balitsky:1987bk, Belyaev:1985wza}:
\begin{align}
\label{edmn13}
S_{q}(x)&= \frac{1}{2 \pi x^2}\Big(i \frac{\xslash}{x^2}- \frac{m_q}{2}\Big) 
-i\frac { g_s }{16 \pi^2 x^2} \int_0^1 dv \, G^{\mu \nu} (vx)
\bigg[\bar v \rlap/{x} 
\sigma_{\mu \nu} + v \sigma_{\mu \nu} \rlap/{x}
 \bigg],\\
% \nonumber\\
%\end{align}%
%\begin{align}
S_{c}(x)&=\frac{m_{c}^{2}}{4 \pi^{2}} \bigg[ \frac{K_{1}\big(m_{c}\sqrt{-x^{2}}\big) }{\sqrt{-x^{2}}}
+i\frac{{\xslash}~K_{2}\big( m_{c}\sqrt{-x^{2}}\big)}
{(\sqrt{-x^{2}})^{2}}\bigg]
%\nonumber\\
%&
-i\frac{m_{c}\,g_{s} }{16\pi ^{2}}  \int_0^1 dv \,G^{\mu \nu}(vx)\bigg[  
    \frac{K_{1}\big( m_{c}\sqrt{-x^{2}}\big) }{\sqrt{-x^{2}}}(\sigma _{\mu \nu }{\xslash}
+{\xslash}\sigma _{\mu \nu })
\nonumber\\
&
 +2\sigma_{\mu \nu }K_{0}\big( m_{c}\sqrt{-x^{2}}\big)\bigg],
 \label{edmn14}
\end{align}%
where $G^{\mu\nu}(x)$ is the gluon field strength tensor,  $v$ is the line variable, and $K_n(m_c\sqrt{-x^2})$ are the Bessel functions. 

The photon's interaction with quarks is incorporated through two distinct mechanisms. The first involves a direct, perturbative coupling to a quark line. In this case, one of the free propagators in Eq.~(\ref{QCD1}) is replaced by its counterpart that includes an electromagnetic interaction vertex:
\begin{align}
\label{free}
S^{\rm free}(x) \rightarrow \int d^4z\, S^{\rm free} (x-z)\,\rlap/{\!A}(z)\, S^{\rm free} (z)\,.
\end{align}
Here, $S^{\rm free}(x)$ corresponds to the leading term of the light- and heavy quark propagators. 

The second mechanism accounts for the non-perturbative coupling of the photon. This is described by substituting one of the light-quark propagators with a matrix element that encapsulates the photon's quark-antiquark structure:
\begin{align}
\label{edmn15}
S_{\mu\nu}^{ab}(x) \rightarrow -\frac{1}{4} \big[\bar{q}^a(x) \Gamma_i q^b(0)\big]\big(\Gamma_i\big)_{\mu\nu},
\end{align}  
where $\Gamma_i = \mathrm{1}, \gamma_5, \gamma_\mu, i\gamma_5 \gamma_\mu, \sigma_{\mu\nu}/2$. This substitution leads to matrix elements of the form $\langle \gamma(q)\vert \bar{q}(x) \Gamma_i q(0) \vert 0\rangle$ and $\langle \gamma(q)\vert \bar{q}(x) \Gamma_i G_{\mu\nu}q(0) \vert 0\rangle$, which are parameterized in terms of the DAs of the photon, following the established formalism~\cite{Ball:2002ps}. 
 By systematically implementing the substitutions from Eqs.~(\ref{free}) and (\ref{edmn15}), both the perturbative and non-perturbative photon interactions are fully incorporated into the QCD representation of the correlation function. The resulting expressions in coordinate space are then transformed to momentum space via a Fourier transformation, yielding the final form used for the subsequent QCD light-cone sum rule analysis.

%\subsection{The sum rules for the electromagnetic multipole moments}
  
Having obtained expressions for the correlation function at both the hadronic and quark-gluon levels, the next step involves deriving the sum rules for the electromagnetic multipole moments. The resulting analytical expressions for the magnetic dipole, electric quadrupole, and magnetic octupole moments of the $P^{\Lambda}_{\psi s}$ state, calculated using five different interpolating currents, are presented in the following equations:
\begin{align}
 \mu_{ J^i_{\mu}}\, \lambda^2_{J^i_{\mu}} \,e^{-\frac{m^2_{J_\mu^i}}{M^2}} & =  \rho_i (M^2,s_0), \\
 \mathcal Q_{ J^i_{\mu}} \, \lambda^2_{J^i_{\mu}} \,e^{-\frac{m^2_{J_\mu^i}}{M^2}} & =  \Delta_i (M^2,s_0), \\
 \mathcal O_{ J^i_{\mu}}\, \lambda^2_{J^i_{\mu}} \, e^{-\frac{m^2_{J_\mu^i}}{M^2}}& =  \Omega_i (M^2,s_0),
\end{align}
\text{with } i = 1,2,\dots,5 \text{ labeling the five interpolating currents, where}

\begin{align}
\rho_i (M^2,s_0) &= \frac{1}{2}\Big[ F_1^{J^i_{\mu}}(M^2,s_0) + F_2^{J^i_{\mu}}(M^2,s_0) \Big], \\
%%%%%%%%%%%%%%%%%%%%%%%%%
\Delta_i (M^2,s_0) &= \frac{1}{2} F_1^{J^i_{\mu}}(M^2,s_0) + m_{J_\mu^i} F_3^{J^i_{\mu}}(M^2,s_0), \\
%%%%%%%%%%%%%%%%%%%%%%
\Omega_i (M^2,s_0) &= \frac{1}{2} \Big[ F_1^{J^i_{\mu}}(M^2,s_0) + F_2^{J^i_{\mu}}(M^2,s_0) \Big]
+ m_{J_\mu^i} F_3^{J^i_{\mu}}(M^2,s_0) + 2 m^2_{J_\mu^i} F_4^{J^i_{\mu}}(M^2,s_0).
\end{align}

Given the extensive and structurally similar nature of the analytical results for the five currents, we provide the explicit expressions only for the first interpolating current, $J_\mu^1$, in the Appendix as a representative example. The corresponding results for the other currents, $J_\mu^2$ to $J_\mu^5$, are obtained analogously.

\end{widetext}

\section{Numerical Illustrations}\label{numerical}

This section is devoted to the numerical investigation of the QCD light-cone sum rule aimed at evaluating the electromagnetic multipole moments of the $P^{\Lambda}_{\psi s}$ state. In order to extract the magnetic dipole moment, several input parameters are required. These include the pentaquark masses ($m_{J^i_{\mu}}$) and residues ($\lambda_{J^i_{\mu}}$), the quark masses ($m_u, m_d, m_s, m_c$), the quark condensates ($\langle \bar q q \rangle$ and $\langle \bar s s \rangle$), the gluon condensate ($\langle g_s^2 G^2 \rangle$), the magnetic susceptibility of the quark condensate $\chi$, and $f_{3\gamma}$—a non-perturbative constant measuring the chirality-flipping soft part of a photon's structure.  The set of numerical inputs adopted in this study is summarized in Table \ref{inputparameter}. For the numerical evaluation, the up and down quark masses are set to zero ($m_u = m_d = 0$), while contributions proportional to $m_s$ are retained. The photon DAs, along with their explicit expressions and the necessary numerical parameters, are adopted from~\cite{Ball:2002ps}. In the present study, the calculations incorporate photon DAs up to twist-4 in order to account consistently for higher-twist effects.
%In the present study, the calculations incorporate photon distribution amplitudes (DAs) up to twist-4, including higher-twist parameters such as the magnetic susceptibility $\chi$ and the nonperturbative constant $f_{3\gamma}$, in order to account consistently for these effects.
   %\begin{widetext}
 \begin{table}[htb!]
	\addtolength{\tabcolsep}{10pt}
	\caption{Input parameters used in the numerical analysis.}
	\label{inputparameter}
	%	\begin{center}
%	\begin{ruledtabular}
		%\scalebox{1.0}{
\begin{tabular}{l|ccc}
               \hline\hline
                \\
Inputs & Values \\
 \\
    \hline\hline
$m_s$&$ 93.5 \pm 0.8$~MeV \cite{ParticleDataGroup:2024cfk}\\
$m_c$&$ 1.273 \pm 0.0046$~GeV \cite{ParticleDataGroup:2024cfk}\\
$m_{J^1_{\mu}}$&   $4.46 \pm 0.10$ GeV~\cite{Wang:2025fqh}\\    
$m_{J^2_{\mu}} $&  $4.42 \pm 0.10$ GeV~\cite{Wang:2025fqh}\\  
$m_{J^3_{\mu}}$&   $4.47 \pm 0.10$ GeV~\cite{Wang:2025fqh}\\ 
$m_{J^4_{\mu}} $&  $4.47 \pm 0.10$ GeV~\cite{Wang:2025fqh}\\  
$m_{J^5_{\mu}}$&   $4.51 \pm 0.10$ GeV~\cite{Wang:2025fqh}\\  
$\lambda_{J^1_{\mu}}  $&$  (1.76 \pm 0.28) \times 10^{-3}~\mbox{GeV}^6$~\cite{Wang:2025fqh}\\ 
$\lambda_{J^2_{\mu}}  $&$  (1.68 \pm 0.27) \times 10^{-3}~\mbox{GeV}^6$~\cite{Wang:2025fqh}\\ 
$\lambda_{J^3_{\mu}}  $&$  (3.05 \pm 0.49) \times 10^{-3}~\mbox{GeV}^6$~\cite{Wang:2025fqh}\\ 
$\lambda_{J^4_{\mu}}  $&$  (3.04 \pm 0.50) \times 10^{-3}~\mbox{GeV}^6$~\cite{Wang:2025fqh}\\    
$\lambda_{J^5_{\mu}}  $&$  (1.87 \pm 0.30) \times 10^{-3}~\mbox{GeV}^6$~\cite{Wang:2025fqh}\\
$\langle \bar qq\rangle $&$ (-0.24 \pm 0.01)^3$~GeV$^3$\cite{Ioffe:2005ym}\\
$\langle \bar ss\rangle $&$ 0.8 \times (-0.24 \pm 0.01)^3~~$~GeV$^3$ \cite{Ioffe:2005ym}\\
$ \langle g_s^2G^2\rangle  $&$ 0.48 \pm 0.14 $~GeV$^4$~\cite{Narison:2018nbv}\\
$f_{3\gamma} $&$ -0.0039 $~GeV$^2$~\cite{Ball:2002ps}\\
$\chi $&$ -2.85 \pm 0.5 $~GeV$^{-2}$~\cite{Rohrwild:2007yt}\\
                                      \hline\hline
 \end{tabular}
%}
%\end{center}
%\end{ruledtabular}
\end{table}
   
%\end{widetext}
Besides the previously mentioned input parameters, two additional quantities are necessary for the numerical evaluation: the continuum threshold, $\mathrm{s_0}$, and the Borel parameter, $\mathrm{M^2}$. In principle, the results should be independent of these parameters, but in practice, some dependence is unavoidable. Therefore, it is essential to determine a working region in which the numerical predictions exhibit minimal sensitivity to variations in $\mathrm{s_0}$ and $\mathrm{M^2}$.  
The working region for the Borel parameter, $\mathrm{M^2}$, defined as the range over which the numerical results show minimal sensitivity to its variation, is determined by the constraints inherent to the adopted methodology. These constraints correspond to the pole contribution (PC) and the convergence of the operator product expansion (CVG). The definitions and quantitative evaluation of these criteria are provided by the following expressions:
\begin{align}
 \mathrm{PC} &=\frac{\rho (\mathrm{M^2},\mathrm{s_0})}{\rho (\mathrm{M^2},\infty)} >35\%, \\
%\nonumber\\
 \mathrm{CVG}&=\frac{\rho^{\mathrm{DimN}} (\mathrm{M^2},\mathrm{s_0})}{\rho (\mathrm{M^2},\mathrm{s_0})}<5\%.
 \end{align}
  
 In this analysis, $\rho_i^{\text{DimN}}(\mathrm{M}^2, s_0)$ denotes the highest-dimensional contributions in the operator product expansion of $\rho_i(\mathrm{M}^2, s_0)$.  Here, only the formalism corresponding to the magnetic dipole moment spectral density, $\rho_i(\mathrm{M}^2, s_0)$, is explicitly presented. Similar analyses have also been performed for the electric quadrupole spectral density, $\Delta_i(\mathrm{M}^2, s_0)$, and for the magnetic quadrupole spectral density, $\Omega_i(\mathrm{M}^2, s_0)$.  On the QCD side, the leading higher-order effects stem from the dimension-7 condensates $\langle g_s^2 G^2 \rangle \langle \bar q q \rangle$ and $\langle g_s^2 G^2 \rangle \langle \bar s s \rangle$. The CVG analysis is therefore performed by including these D7 terms, and the parameters reported in Table~\ref{parameter} are obtained within this setup. The QCD representation further involves several operator structures of lower dimensionalities, such as $\langle g_s^2 G^2 \rangle f_{3\gamma}$ (D6), $\langle g_s^2 G^2 \rangle \langle \bar q q \rangle \chi$ and $\langle g_s^2 G^2 \rangle \langle \bar s s \rangle \chi$ (D5), $\langle \bar q q \rangle$ and $\langle \bar s s \rangle$ (D3), $f_{3\gamma}$ (D2), and $\langle \bar q q \rangle \chi$ and $\langle \bar s s \rangle \chi$ (D1).  
 Once these conditions were satisfied, we proceeded with confidence in the robustness of our predictions. Figure~\ref{Msqfig} presents the results specifically for the magnetic dipole moment, demonstrating that within the selected Borel window, the PC remains substantially larger than the continuum contribution, ensuring the dominance of the ground state. Simultaneously, the relative sizes of the condensate terms indicate a satisfactory CVG. For completeness, Fig.~\ref{Msqfig} also shows the dependence of the $P^{\Lambda}_{\psi s}$ magnetic dipole moment on the Borel parameter $\mathrm{M^2}$ and the continuum threshold $\mathrm{s_0}$. As expected, only mild variations are observed within the chosen ranges, although a residual sensitivity to these parameters persists, contributing to the overall uncertainty.
  \begin{widetext}
 \begin{table}[htb!]
	\addtolength{\tabcolsep}{10pt}
	\caption{Working intervals of $\mathrm{s_0}$ and $\mathrm{M^2}$ together with the PC and CVG results.}
	\label{parameter}
		\begin{center}
		%\begin{ruledtabular}
\begin{tabular}{l|ccccc}
	   \hline\hline
	   \\
   State& $\mathrm{s_0}$~\mbox{(GeV$^2$)}&$\mathrm{M^2}$~\mbox{(GeV$^2$)}&\mbox{CVG} $(\%)$&\mbox{PC} $(\%)$\\
   \\
\hline\hline
%\\
$J_\mu^1$ &$ [26.0, 28.0]$   & $ [2.5, 3.1] $ & [0.10, 0.12] &[66.18, 42.54]\\
%\\
$J_\mu^2$ &$ [26.0, 28.0]$   &  $ [2.7, 3.3]$ & [0.09, 0.13] &[68.18, 47.20]\\
%\\
$J_\mu^3$ &$ [26.0, 28.0]$   &  $ [2.5, 3.1]$ & [0.10, 0.13] &[63.88, 40.45]\\
%\\
$J_\mu^4$ &$ [26.0, 28.0]$   &  $ [2.5, 3.1]$ & [0.15, 0.18] &[67.37, 43.18]\\
%\\
$J_\mu^5$ &$ [26.0, 28.0]$   &  $ [2.5, 3.1]$ & [0.10, 0.14] &[67.60, 43.83]\\
%\\
	   \hline\hline
\end{tabular}
\end{center}
%\end{ruledtabular}
\end{table}
\end{widetext}

With all the necessary input parameters and auxiliary quantities, such as the continuum threshold $\mathrm{s_0}$ and the Borel parameter $\mathrm{M^2}$, determined, we now present the resulting numerical values. The complete set of results, including the estimated uncertainties arising from the intrinsic variability of the input parameters, is summarized in Table~\ref{sonuc}.
%  \begin{widetext}
 \begin{table}[htp]
	\addtolength{\tabcolsep}{6pt}
	\caption{	Numerical results of the magnetic dipole (in units of $\mu_N$), electric quadrupole (in units of  fm$^2$) and magnetic octupole (in units of  fm$^3$) moments of the $P^{\Lambda}_{\psi s}$.}
	\label{sonuc}
		\begin{center}
		%\begin{ruledtabular}
\begin{tabular}{l|ccccc}
	   \hline\hline
	   \\
   State& $\mu_{P^{\Lambda}_{\psi s}}$& $\mathcal Q_{P^{\Lambda}_{\psi s}} (\times 10^{-2}) $ &$\mathcal O_{P^{\Lambda}_{\psi s}} (\times 10^{-2}) $\\
   \\
\hline\hline
%\\
$J_\mu^1$ & $ -2.28 \pm 0.57$  &$ -0.90 \pm 0.18$    & $-0.49 \pm 0.10 $\\
%\\
$J_\mu^2$ & $ -0.15  \pm 0.04$ &$  -1.44 \pm 0.29$   &  $-1.02 \pm 0.21$\\
%\\
$J_\mu^3$ & $ ~~1.60 \pm 0.40$ & $~~0.24 \pm 0.05 $  &  $ -0.13 \pm 0.03 $\\
%\\
$J_\mu^4$ & $ -1.26 \pm 0.32$  &$ ~3.17 \pm 0.64$    &  $ ~~1.44 \pm 0.30 $\\
%\\
$J_\mu^5$ & $ ~~3.36 \pm 0.84$ &$ -2.32 \pm 0.47$    &  $ -1.88 \pm 0.38 $\\
%\\
	   \hline\hline
\end{tabular}
\end{center}
%\end{ruledtabular}
\end{table}
%\end{widetext}
 %\end{widetext}

 As presented in Table~\ref{sonuc}, the magnetic dipole moments of the $P^{\Lambda}_{\psi s}$ pentaquark, evaluated using several distinct interpolating currents but sharing the same quark content, display noticeable variations. This behavior may indicate the presence of more than one $P^{\Lambda}_{\psi s}$-type pentaquark state (see the discussion following Eq.~(\ref{mqo2})), characterized by identical quantum numbers and similar quark configurations, yet differing in their electromagnetic properties. As discussed earlier, the employed interpolating currents share the same quantum numbers and consequently lead to nearly degenerate mass predictions for the $P^{\Lambda}_{\psi s}$ states~\cite{Wang:2025fqh}. Nevertheless, the extracted magnetic dipole, electric quadrupole and magnetic octupole moments exhibit a pronounced dependence on the internal diquark--diquark--antiquark organization and the microscopic structure of the considered state. Although it is often assumed and expected that a change in the hadronic basis does not alter physical observables, this assumption may not hold for quantities directly linked to internal dynamics, such as magnetic dipole moments. Since these observables are highly sensitive to the spatial and spin configurations of the constituent quarks, a transformation of the interpolating basis effectively modifies the internal structure of the state, which can, in turn, lead to substantial variations in the computed electromagnetic multipole moments. Indeed, several studies~\cite{Ozdem:2024txt,Ozdem:2024dbq,Azizi:2023gzv,Ozdem:2024rqx,Ozdem:2022iqk} employing different types of interpolating currents have reported significant differences in the magnetic dipole moments of multi-quark systems. These findings collectively suggest that variations in the choice of interpolating currents, as well as in the isospin or charge configurations of the investigated states, can yield distinct magnetic dipole moments even for hadrons with the same quark content.

 As seen from Table~\ref{sonuc}, the higher electromagnetic multipole moments, similar to the magnetic dipole moment, exhibit a strong dependence on the specific diquark–diquark–antiquark configuration adopted for the $P^{\Lambda}_{\psi s}$ pentaquark. The magnitudes of the electric quadrupole and magnetic octupole moments are considerably smaller than that of the magnetic dipole moment, yet their nonvanishing values clearly indicate a nonspherical charge distribution. The relative magnitudes and signs of these higher moments provide insight into the deformation and orientation of the underlying hadronic structure.  For the interpolating currents $J_\mu^1$, $J_\mu^2$, and $J_\mu^5$, both the electric quadrupole and magnetic octupole moments are negative, suggesting an oblate deformation, where the charge distribution is flattened along the symmetry axis. In contrast, the $J_\mu^3$ and $J_\mu^4$ currents yield positive values for the quadrupole moments—implying a prolate deformation, in which the charge distribution is elongated along the symmetry axis. Moreover, for $J_\mu^3$, the opposite signs of the quadrupole and octupole moments (positive and negative, respectively) indicate that the charge and geometric distributions are oriented differently. These results suggest that the multipole structure of the $P^{\Lambda}_{\psi s}$ pentaquark is highly sensitive to the underlying diquark configuration encoded in the interpolating current. Such sensitivity may provide a useful probe for distinguishing between different internal diquark--diquark--antiquark configurations in future experimental or lattice studies.

To gain a deeper understanding of the observed variations in the results, we have analyzed the contributions of individual quarks, with the corresponding findings summarized in Tables~\ref{parameter2}--\ref{parameter4}. This can be attained by choosing the relevant charge factors $e_u$, $e_d$, $e_s$, and $e_c$.  These results are evaluated using the central values of all input parameters. 
The individual quark contributions to the magnetic dipole, electric quadrupole, and magnetic octupole moments of the $P^{\Lambda}_{\psi s}$ pentaquark reveal several interesting features. For the magnetic dipole moment, the up and down quarks provide the dominant contributions, often with opposite signs that partially cancel each other. The strange and charm quarks generally contribute less, although their effects become more pronounced in certain interpolating currents, such as $J_\mu^2$ and $J_\mu^3$. 
In the case of the electric quadrupole moments, the charm quark contributions are substantial in most currents, highlighting its influence on the deformation of the charge distribution. Notably, the $J_\mu^4$ current exhibits large positive contributions from all quarks, leading to a strongly positive total quadrupole moment, consistent with a prolate deformation. Conversely, for $J_\mu^1$, $J_\mu^2$, and $J_\mu^5$, the combined contributions produce negative total quadrupole moments, indicative of an oblate geometry. 
The magnetic octupole moments show similar trends. The up and down quarks largely dominate, with partial cancellations shaping the total value, while the strange and charm quarks provide moderate corrections. The interplay of contributions from all quarks results in both positive and negative total octupole moments across different interpolating currents, reflecting the sensitivity of the octupole structure to the internal quark configuration.  
Overall, these findings underline that the electromagnetic multipole moments of the $P^{\Lambda}_{\psi s}$ are highly sensitive to the underlying quark arrangement, and different interpolating currents probe distinct structural aspects of the pentaquark. 
%Furthermore, to illustrate how individual quark contributions influence the deformations observed in the electric quadrupole and magnetic octupole moments, we present these effects explicitly in Figs.~\ref{quark_deformation11}--\ref{quark_deformation12}. These figures clearly indicate the role of each quark in modifying the multipole distributions, allowing one to discern which quark contributes to specific changes in the overall structure.

To dissect the microscopic mechanisms behind the structural deformations, we decompose the electric quadrupole and magnetic octupole moments into individual quark contributions, as shown in Figs.~\ref{quark_deformation11}--\ref{quark_deformation12}. This decomposition uncovers several critical insights. First, regarding the quadrupole deformation, the charm quark emerges as the principal actor. Its substantial negative contribution in oblate configurations (e.g., $J_{\mu}^{1}$, $J_{\mu}^{2}$, $J_{\mu}^{5}$) and positive contribution in prolate ones (e.g., $J_{\mu}^{3}$, $J_{\mu}^{4}$) indicate that the spatial distribution of the heavy charm quark is a major factor in the system's elongation or flattening. This dominance can be understood by considering the large mass of the charm quark, which, in the spirit of heavy-quark effective theory, suggests that the lighter degrees of freedom adjust their dynamics around it, thereby making the charm quark's distribution a primary driver of the collective deformation. 
Second, we observe a complex interplay for the magnetic octupole moment. While light quarks provide the largest individual contributions, they frequently exhibit partial cancellation (e.g., the up and down quark contributions often oppose each other). The net octupole moment is then often determined by the smaller, yet uncompensated, contributions from the strange and charm quarks. This reveals that the higher-order magnetic structure is exceptionally sensitive to minor asymmetries in the strange-charm sector, acting as a magnifying glass for subtle flavor-dependent dynamics. 
Collectively, these results provide a structural fingerprint for each configuration. For example, a future experimental observation of a large oblate deformation would strongly suggest a configuration where the charm quark's wavefunction is concentrated in the equatorial plane, as seen in our $J_{\mu}^{5}$ current. Most importantly, this flavor decomposition maps abstract multipole moments directly onto specific quark-level behaviors, transforming them from gross structural indicators into precise diagnostic tools for benchmarking theoretical models against future experimental data.

\begin{table}[htp]
\addtolength{\tabcolsep}{6pt}
\caption{Individual quark contributions to the magnetic dipole moments (in units of $\mu_N$).}
	\label{parameter2}
		\begin{center}
		%\begin{ruledtabular}
\begin{tabular}{l|ccccc}
	   \hline\hline
	   \\
   State& $\mu_u$& $\mu_d$&$\mu_s$&$\mu_c$&$\mu_{tot}$\\
   \\
\hline\hline
%\\
$J_\mu^1$ &$ -3.52$  &$ ~~1.76$ &  $\sim 0$ & $-0.52$  &$-2.28$\\
%\\
$J_\mu^2$ & $ -3.62$ &$ ~~1.81$ &  $~~3.62$ & $-1.96$  &$-0.15$\\
%\\
$J_\mu^3$ & $-8.82$  &$~~4.41$  &  $~~4.71$ & $~~1.30$ &$~~1.60$\\
%\\
$J_\mu^4$ & $-9.34$  &$ ~~2.24$ &  $~~3.94$ & $~~1.90$ &$-1.26$\\
%\\
$J_\mu^5$ & $\sim 0$ &$\sim 0$  &  $~~3.80$ & $-0.44$  &$~~3.36$\\
%\\
	   \hline\hline
\end{tabular}
\end{center}
%\end{ruledtabular}
\end{table}

%\end{widetext}
%   \begin{widetext}
   
\begin{table}[htp]
\addtolength{\tabcolsep}{6pt}
\caption{Individual quark contributions to the electric quadrupole moments ($\times 10^{-2}$ in units of fm$^2$).}
	\label{parameter3}
		\begin{center}
		%\begin{ruledtabular}
\begin{tabular}{l|ccccc}
	   \hline\hline
	   \\
   State& ${\mathcal Q}_u$& ${\mathcal Q}_d$&${\mathcal Q}_s$&${\mathcal Q}_c$&${\mathcal Q}_{tot}$\\
   \\
\hline\hline
%\\
$J_\mu^1$ &$ ~~0.84$  &$-0.42$   &  $\sim 0$ & $-1.32$   &$-0.90$\\
%\\
$J_\mu^2$ & $~~0.88$  &$-0.44$   &  $-0.80$  & $~~1.00$  &$-1.44$\\
%\\
$J_\mu^3$ & $~~0.06$  &$-0.03$   &  $-1.10$  & $~~1.31$  &$~~0.24$\\
%\\
$J_\mu^4$ & $~~0.06$  &$ ~~0.55$ &  $~~1.18$ & $~~1.38$  &$~~3.17$\\
%\\
$J_\mu^5$ & $\sim 0$  &$\sim 0$  &  $-0.89$  & $-1.43$   &$-2.32$\\
%\\
\hline\hline
\end{tabular} 
\end{center}
%\end{ruledtabular}
\end{table}

%\end{widetext}
%   \begin{widetext}
   
\begin{table}[htp]
\addtolength{\tabcolsep}{6pt}
\caption{Individual quark contributions to magnetic octupole moments ($\times 10^{-2}$ in units of fm$^3$).}
	\label{parameter4}
		\begin{center}
		%\begin{ruledtabular}
\begin{tabular}{l|ccccc}
	   \hline\hline
	   \\
State& ${\mathcal O}_u$& ${\mathcal O}_d$&${\mathcal O}_s$&${\mathcal O}_c$&${\mathcal O}_{tot}$\\
   \\
\hline\hline
%\\
$J_\mu^1$ &$~~0.82$   &$ -0.41$  &  $\sim 0$ & $-0.90$  &$-0.49$\\
%\\
$J_\mu^2$ & $ ~~0.86$ &$ -0.43$  &  $-0.86$  & $-0.59$  &$-1.02$\\
%\\
$J_\mu^3$ & $~~1.52$  &$-0.76$   &  $-1.11$  & $~~0.22$ &$-0.13$\\
%\\
$J_\mu^4$ & $~~1.52$  &$ -0.18$  &  $-0.27$  & $~~0.37$ &$~~1.44$\\
%\\
$J_\mu^5$ & $\sim 0$  &$\sim 0$  &  $-0.88$  & $-1.00$  &$-1.88$\\
%\\
	   \hline\hline
\end{tabular} 
\end{center}
%\end{ruledtabular}
\end{table}

%\end{widetext}

A critical test for the structural interpretations of the $P^{\Lambda}_{\psi s}$ state is provided by comparing its predicted magnetic dipole moment across different theoretical frameworks, all assuming $J^P = \frac{3}{2}^-$. Our results, summarized in Table~\ref{sonuc} for the compact pentaquark picture, reveal a spectrum of values that is both wide and informative. 
This diversity stands in stark contrast to existing literature. Calculations based on a molecular configuration yield values of $\mu_{P^{\Lambda}_{\psi s}} = 0.465~\mu_N$ (quark model)~\cite{Li:2021ryu},  $\mu_{P^{\Lambda}_{\psi s}} = -0.231~\mu_N$~\cite{Gao:2021hmv} (quark model) and $\mu_{P^{\Lambda}_{\psi s}} = -1.67 \pm 0.58~\mu_N$ (QCD light-cone sum rules)~\cite{Ozdem:2022kei}. Notably, our predictions for the compact state not only differ in magnitude from these molecular results but also span both positive and negative signs, a direct consequence of the distinct interpolating currents employed. This pronounced model-dependence is, in fact, a key strength of the magnetic moment as an observable. The large variation in our predictions for different currents means that a future experimental measurement of the magnetic moment will not merely provide a number, but will act as a powerful discriminator. It will allow us to rule out entire classes of interpolating currents and, by extension, the specific internal configurations they represent, thereby providing a stringent test for our theoretical understanding of the pentaquark's structure. 
The discrepancies between our compact pentaquark results and the molecular predictions from other studies highlight the pivotal role of the magnetic moment in discriminating between core structural models. The underlying reasons for the spread are multifaceted, stemming from fundamental differences in the assumed spatial organization of quarks (compact vs. loose molecule), the choice of dynamical framework (QCD sum rules vs. quark models), and the associated parameterizations. 
In conclusion, while the precise value remains model-dependent, our analysis firmly establishes the magnetic dipole moment as a powerful diagnostic tool. The substantial differences observed between various structural assumptions suggest that future, more precise experimental constraints on this quantity could be decisive in elucidating the true nature of the $P^{\Lambda}_{\psi s}$ pentaquark.

\section{Structural Interpretation and Experimental Signatures}
\label{sec:discussion}

The pronounced variations in the electromagnetic moments across the five interpolating currents, despite their shared quantum numbers $J^P = \frac{3}{2}^{-}$, indicate a high sensitivity to the internal quark-gluon organization. This diversity suggests that the $P^{\Lambda}_{\psi s}$ spectrum may host multiple near-degenerate states or that a single state admits several distinct structural configurations. Below, we interpret our results and provide specific experimental signatures to discriminate between these possibilities.

\subsection{Structural Interpretation of the Currents}

The five currents employed in this study can be grouped into distinct structural categories based on their diquark content and Lorentz architecture:

\begin{itemize}
    \item $J^3_\mu$ and $J^4_\mu$: Compact Axial-Vector Dominated Configurations. These currents, built from two axial-vector diquarks, are optimized to probe compact multiquark cores. The positive quadrupole moment of $J^4_\mu$ ($\mathcal{Q} = 3.17$) signals a prolate charge distribution, characteristic of a tightly-bound, anisotropic system. The significant magnetic moments arise from the aligned spins in the axial-vector diquarks.

    \item $J^1_\mu$ and $J^2_\mu$: Scalar-diquark dominated configurations.  These currents contain substantial scalar diquark components ($C\gamma_5$), which are energetically favored in one-gluon exchange and couple strongly to compact configurations. However, through Fierz rearrangement, the same scalar diquark structure may also overlap with molecular-type components such as $\bar{D}^{(*)}\Xi_c$, where the light-quark subsystem forms a scalar correlation. The consistently negative quadrupole moments ($\mathcal{Q} < 0$) indicate an oblate deformation, consistent with either compact scalar clustering or peripheral charge distributions in loosely bound molecular configurations.

    \item $J^5_\mu$: Sector-Separated Cluster. This current uniquely isolates a non-strange scalar diquark $[ud]$ from an axial-vector strange-charm diquark $[sc]$. This clear flavor-spin separation leads to the largest predicted magnetic moment ($\mu = 3.36~\mu_N$) and a strong oblate deformation ($\mathcal{Q} = -2.32$), serving as a unique fingerprint for this specific clustering pattern.
\end{itemize}

\begin{widetext}
 
\begin{table}[htp]
\centering
\caption{Theoretical benchmarks for experimental discrimination of the $P^{\Lambda}_{\psi s}$ structure. The photon angular distribution, $W(\theta_\gamma)$, refers to the $P^{\Lambda}_{\psi s} \to J/\psi \Lambda \gamma$ decay in the $P^{\Lambda}_{\psi s}$ rest frame, where $\theta_\gamma$ is the angle between the photon momentum and the spin-quantization axis of the $P^{\Lambda}_{\psi s}$.}
\label{table:discriminators}
\begin{tabular}{lccc}
\hline\hline
\\
Current & $\mu$ ($\mu_N$) & $\mathcal{Q}$ ($10^{-2}$ fm$^2$) & Experimental Signature \\
\\
\hline\hline
\\
$J^1_\mu$ & $-2.28 \pm 0.57$ & $-0.90 \pm 0.18$ & Moderate rate. Broad, sin$^2\theta_\gamma$-dominated distribution with oblate distortion. \\
\\
$J^2_\mu$ & $-0.15 \pm 0.04$ & $-1.44 \pm 0.29$ & Highly suppressed rate. Strongly cos$^2\theta_\gamma$-dominated distribution. \\
\\
$J^3_\mu$ & $~~1.60 \pm 0.40$ & $~~0.24 \pm 0.05$ & Significant rate. Nearly pure dipolar (sin$^2\theta_\gamma$) distribution. \\
\\
$J^4_\mu$ & $-1.26 \pm 0.32$ & $~~3.17 \pm 0.64$ & Significant rate. Strongly cos$^2\theta_\gamma$-dominated, sharp peak at poles. \\
\\
$J^5_\mu$ & $~~3.36 \pm 0.84$ & $-2.32 \pm 0.47$ & Largest rate. sin$^2\theta_\gamma$ base with significant equatorial enhancement. \\
\\
\hline\hline
\end{tabular}
\end{table}

\end{widetext}

\subsection{Experimental Discriminators and Future Measurements}

The most direct experimental probe of the electromagnetic structure of the $P^{\Lambda}_{\psi s}$ states is through their radiative decays, accessible in processes like $\Xi_b^- \to J/\psi \Lambda K^-\gamma$ at LHCb and Belle II. The key observables are the radiative decay width (proportional to $|G_M(0)|^2$) and the photon angular distribution, which is highly sensitive to the ratio $G_Q/G_M$.

In Table~\ref{table:discriminators}, we provide a set of theoretical benchmarks that directly connect our predictions to measurable quantities. The angular distribution of the photon in the $P^{\Lambda}_{\psi s}$ rest frame, $W(\theta_\gamma)$, can be parameterized in terms of the multipole ratios, providing a direct experimental handle on the underlying deformation.

The following are decisive experimental scenarios:
\begin{itemize}
    \item A measurement of a large radiative branching fraction combined with a photon angular distribution strongly peaked along the spin axis (large positive $G_Q/G_M$) would be a smoking gun for the prolate, compact configuration of $J^4_\mu$.

    \item Conversely, a large magnetic moment ($\mu \gtrsim 3 \mu_N$) together with a flattened, oblate angular distribution would provide compelling evidence for the sector-separated structure of $J^5_\mu$.

    \item A highly suppressed radiative decay would point towards the $J^2_\mu$-like configuration, where the internal currents largely cancel.
\end{itemize}

Although the extraction of the multipole moments is experimentally challenging, the predicted angular and polarization patterns can, in principle, be resolved in high-statistics analyses of $\Xi_b^- \to J/\psi\Lambda K^-\gamma$ decays. Such analyses are within reach of current amplitude reconstruction techniques at LHCb.  The correlations between the photon emission angle and the spin alignment of the $P^{\Lambda}_{\psi s}$, as established in this study, provide a framework that is largely independent of the specific current choice. Therefore, the present results provide a realistic theoretical basis for future LHCb and Belle II measurements aimed at mapping the electromagnetic geometry of hidden-charm strange pentaquarks.

In addition to experimental efforts, lattice QCD provides a complementary and first-principles framework for investigating the electromagnetic structure of multiquark systems. Although current lattice simulations of hidden-charm pentaquarks remain challenging due to their high masses and multi-hadron nature, recent progress in computing electromagnetic form factors of doubly charmed baryons~\cite{Can:2013zpa, Can:2013tna} and doubly bottom tetraquark \cite{Vujmilovic:2025czt} suggests that analogous calculations for pentaquarks may soon become feasible. Such lattice studies would offer a valuable cross-check of our predictions for the magnetic dipole, electric quadrupole, and magnetic octupole moments, thereby providing an ab initio benchmark for the nonperturbative dynamics captured by QCD sum rules.

\section{Summary and Conclusions}\label{summary}

In this work, we have performed a comprehensive analysis of the electromagnetic properties of the hidden-charm strange pentaquark with $J^{P}=3/2^{-}$ within the framework of QCD light-cone sum rules. Employing five distinct diquark-diquark-antiquark interpolating currents, we have calculated the magnetic dipole, electric quadrupole, and magnetic octupole moments to probe the internal structure of this exotic state.

Our main findings can be summarized as follows:

\begin{itemize}
    \item The electromagnetic multipole moments exhibit strong sensitivity to the choice of interpolating current, demonstrating their intimate connection to the internal quark-gluon structure of the pentaquark.
    
    \item Magnetic dipole moments range from $-2.28\,\mu_{N}$ to $+3.36\,\mu_{N}$, establishing $\mu$ as a powerful discriminator among different internal configurations with identical quantum numbers.
    
    \item Non-zero electric quadrupole and magnetic octupole moments indicate a non-spherical charge distribution, with the sign and magnitude of these moments revealing a complex pattern of prolate and oblate deformations across different currents.
    
    \item Flavor decomposition shows that light quarks dominate the magnetic dipole response, often with significant cancellations, whereas the charm quark significantly influences the quadrupole deformation.

    \item Future studies employing currents explicitly constructed for molecular configurations (e.g., $\bar{D}^{(*)}\Xi_c$ meson-baryon pairs) would provide a direct comparison with the present compact-diquark based results, thereby offering a more definitive test for the internal structure of the $P^{\Lambda}_{\psi s}$ states. 
\end{itemize}

The systematic variation across five distinct currents provides a structural map of the pentaquark configuration space. As detailed in Section~\ref{sec:discussion}, we have provided specific structural interpretations for each current and established experimental benchmarks (Table~\ref{table:discriminators}) that connect our theoretical predictions to measurable quantities in radiative decays. 
Our predictions provide essential benchmarks for future experimental investigations---such as analyses of $\Xi_b^{-} \to J/\psi \Lambda K^{-}\gamma$ at LHCb and Belle II---as well as for complementary lattice QCD studies. These efforts will be crucial for unveiling the true nature of these enigmatic pentaquark states and guiding the interpretation of forthcoming experimental data.  
This multi-current approach establishes a new paradigm for exotic hadron spectroscopy, moving beyond mass-based classification to structure-sensitive observables that directly probe QCD's non-perturbative dynamics.

The present framework can be systematically extended to investigate the electromagnetic properties of non-strange hidden-charm pentaquarks, as well as those with $J^{P} = \frac{1}{2}^{-}$, by constructing appropriate interpolating currents and repeating the analysis outlined in this work. Such studies would provide a more complete spectroscopic map and further test the structural sensitivity of multipole moments.

%\newpage

\begin{widetext}

  \begin{figure}[htp]
\centering
\includegraphics[width=0.3\textwidth]{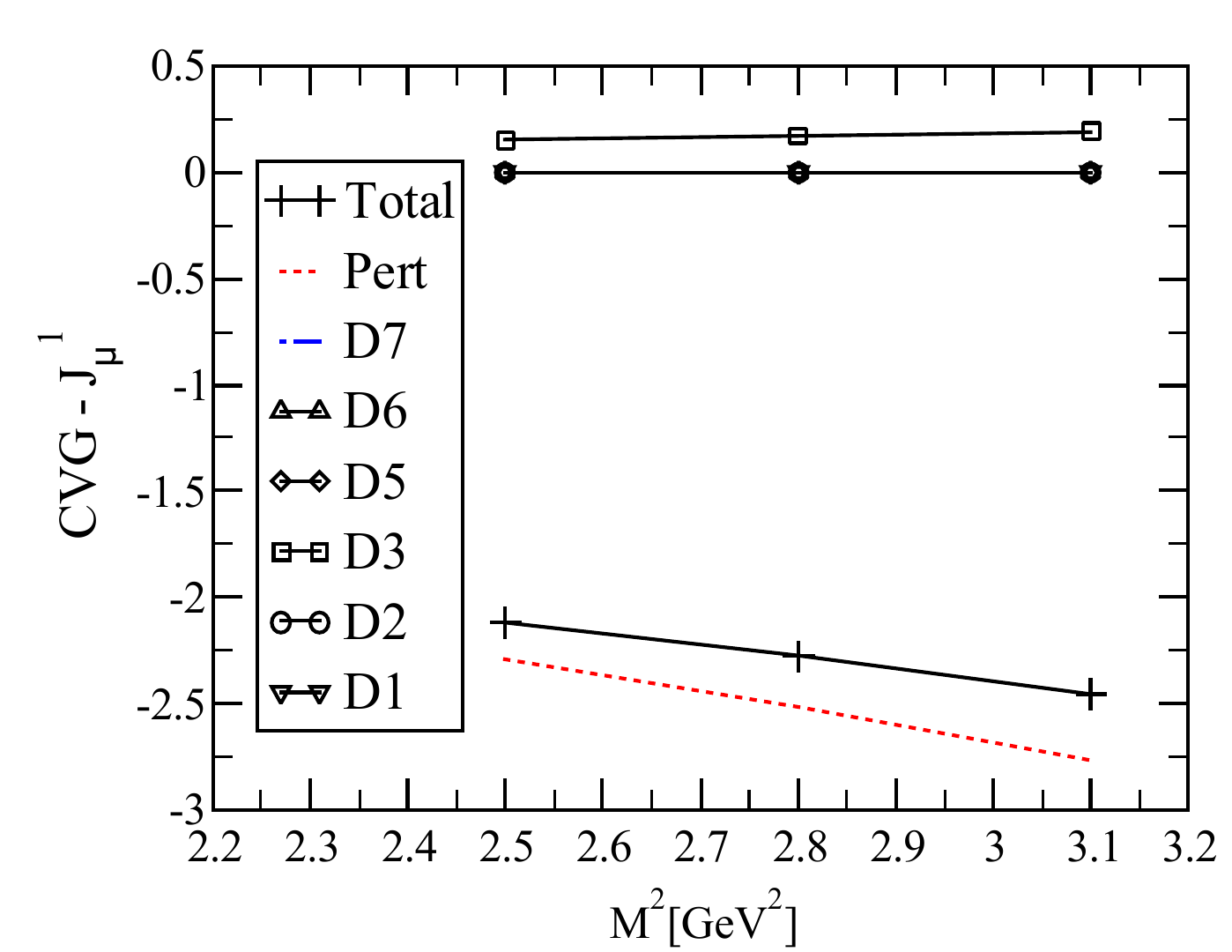}~~ ~~
\includegraphics[width=0.3\textwidth]{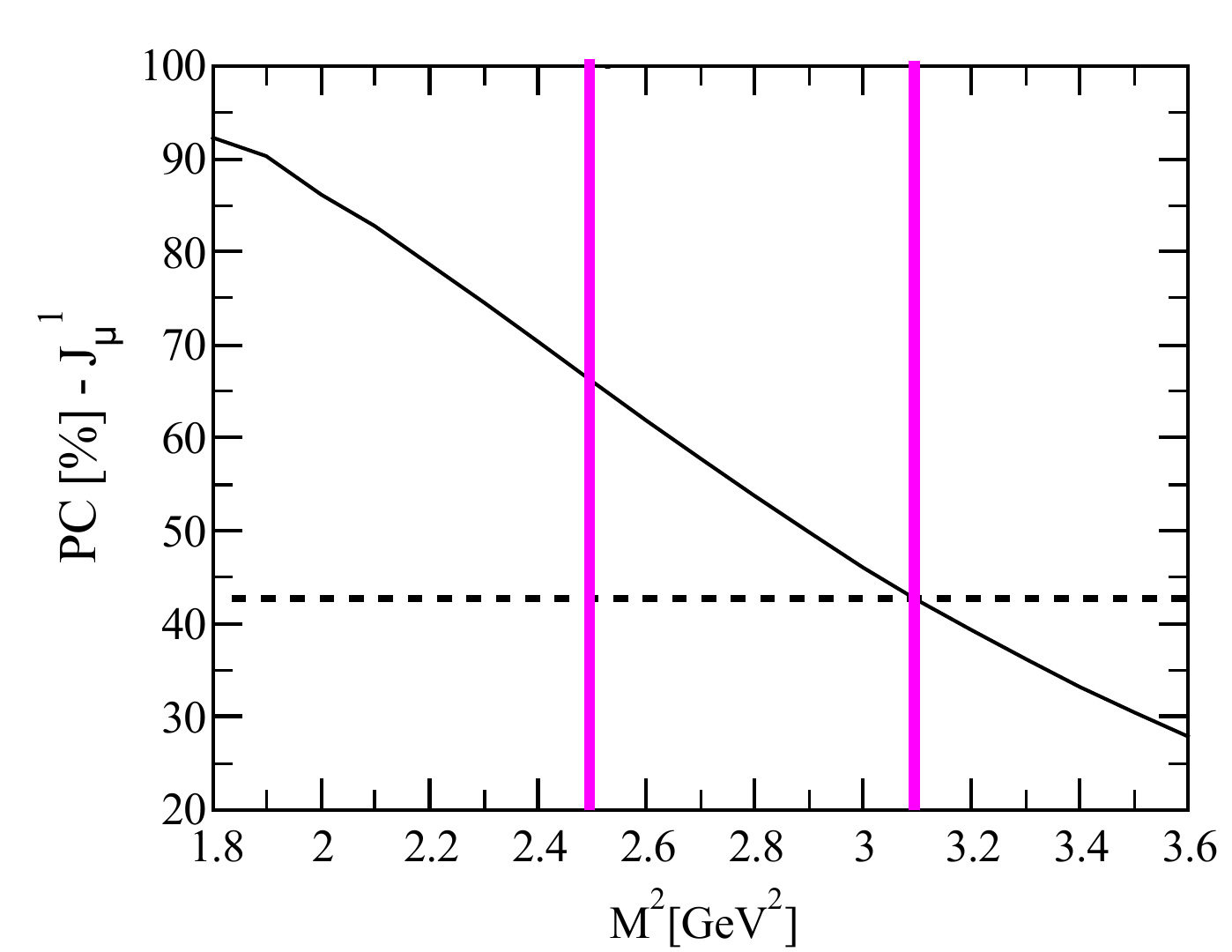}~~ ~~
\includegraphics[width=0.3\textwidth]{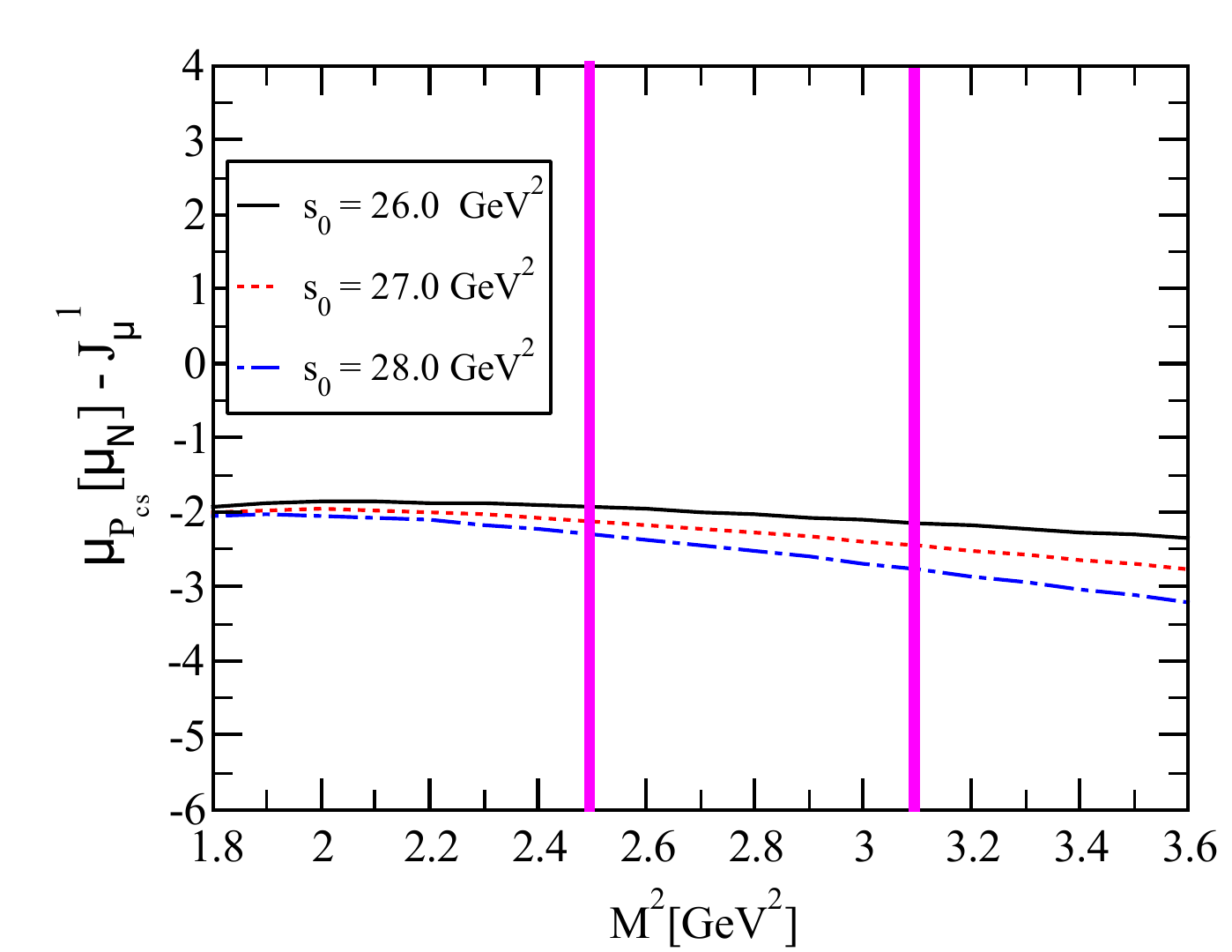}\\

\includegraphics[width=0.3\textwidth]{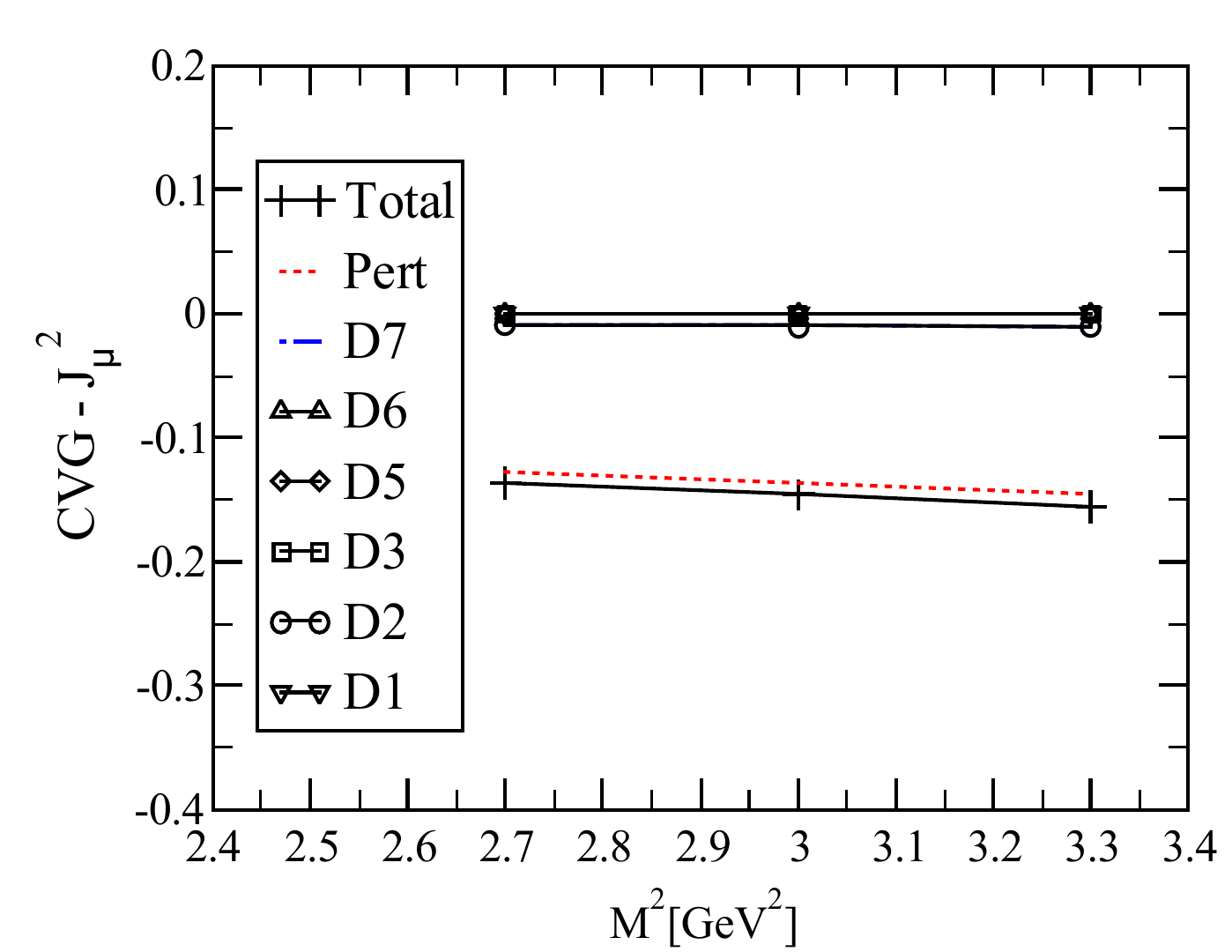}~~ ~~
\includegraphics[width=0.3\textwidth]{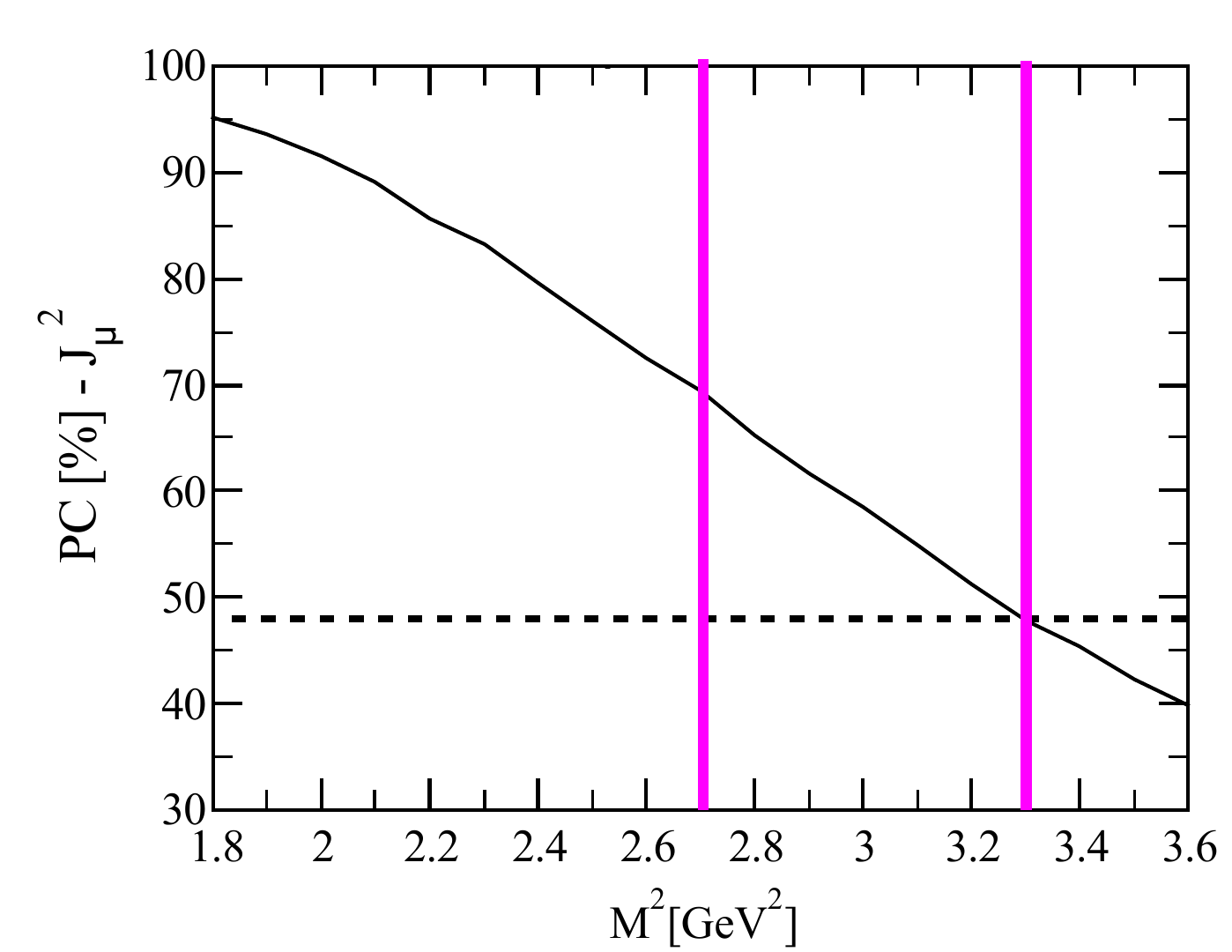}~~ ~~
\includegraphics[width=0.3\textwidth]{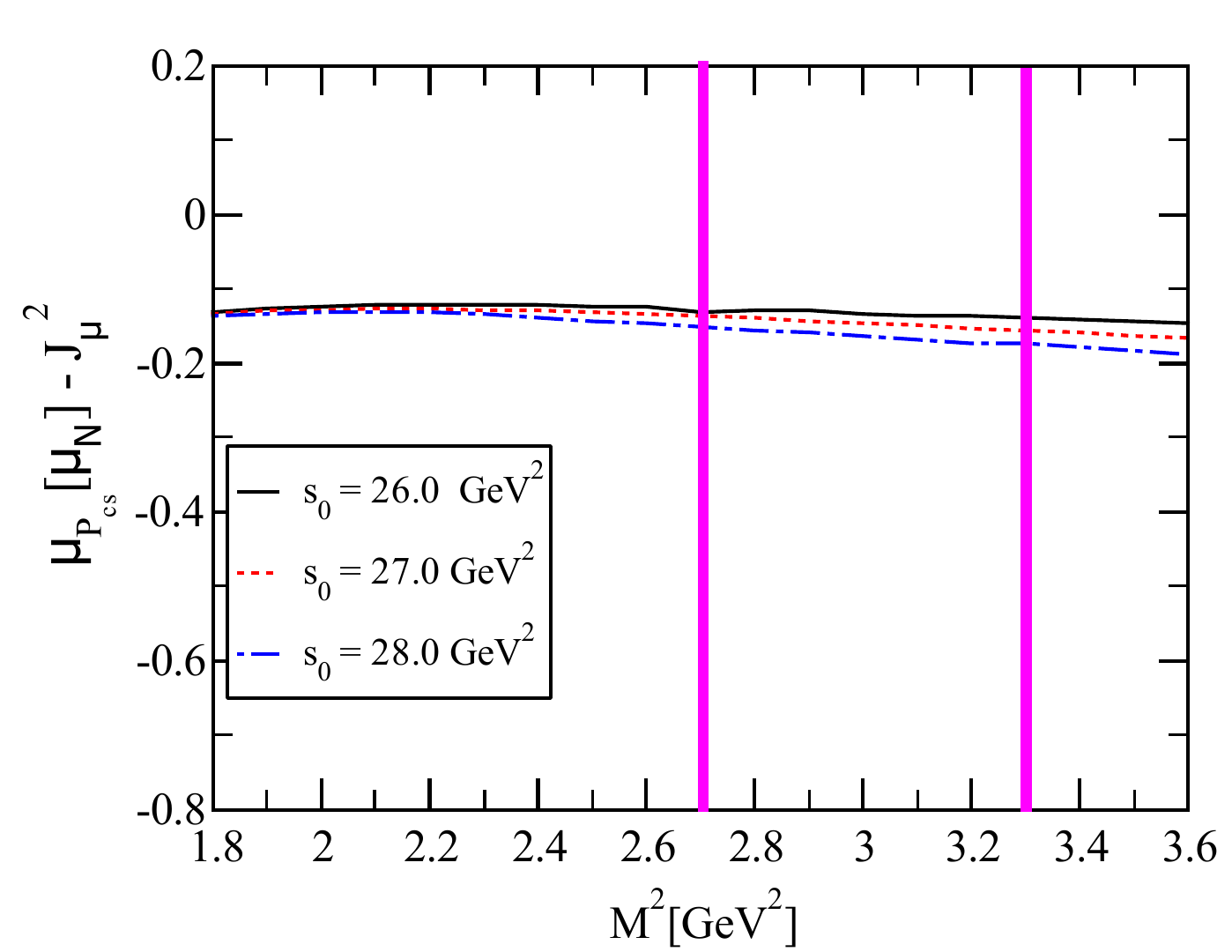}\\

\includegraphics[width=0.3\textwidth]{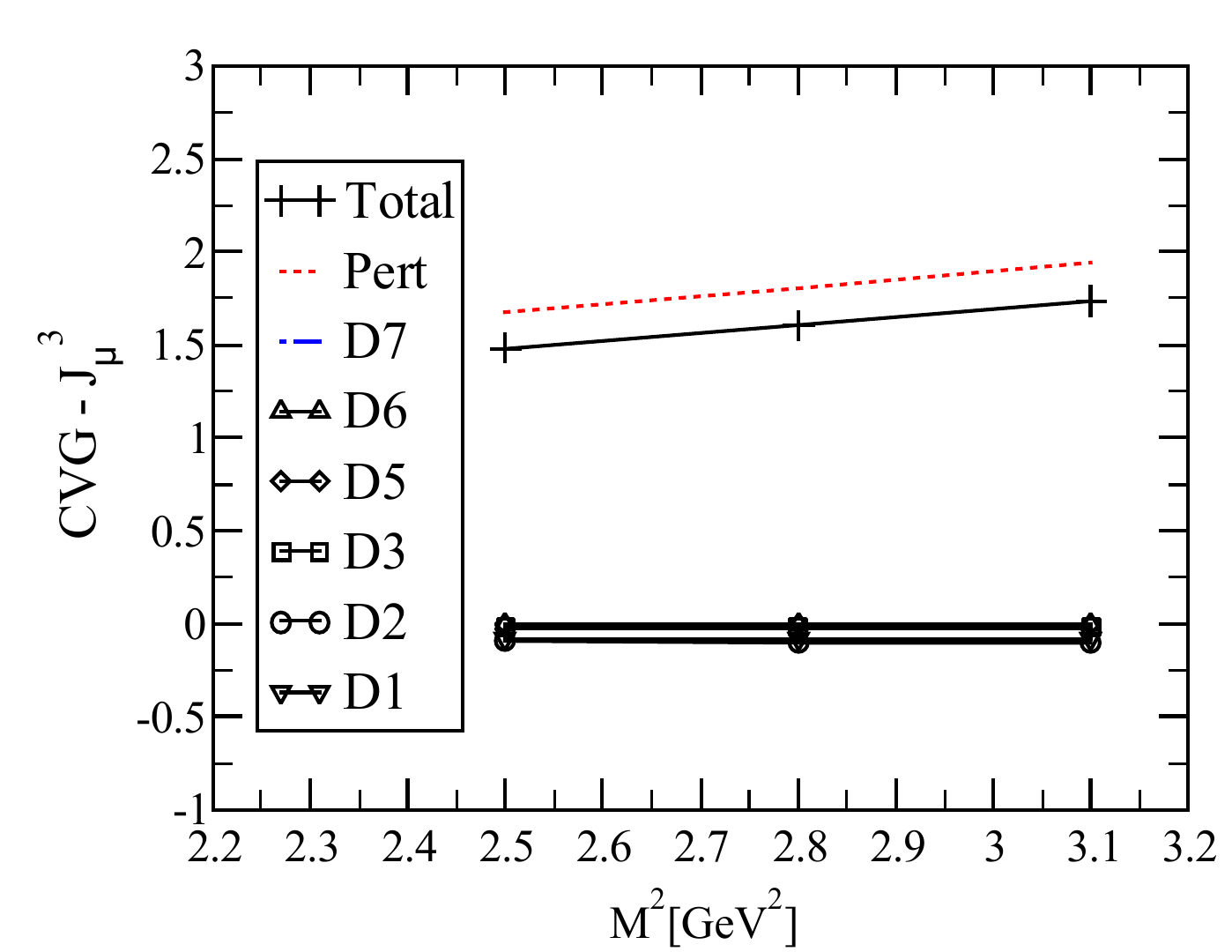}~~ ~~
\includegraphics[width=0.3\textwidth]{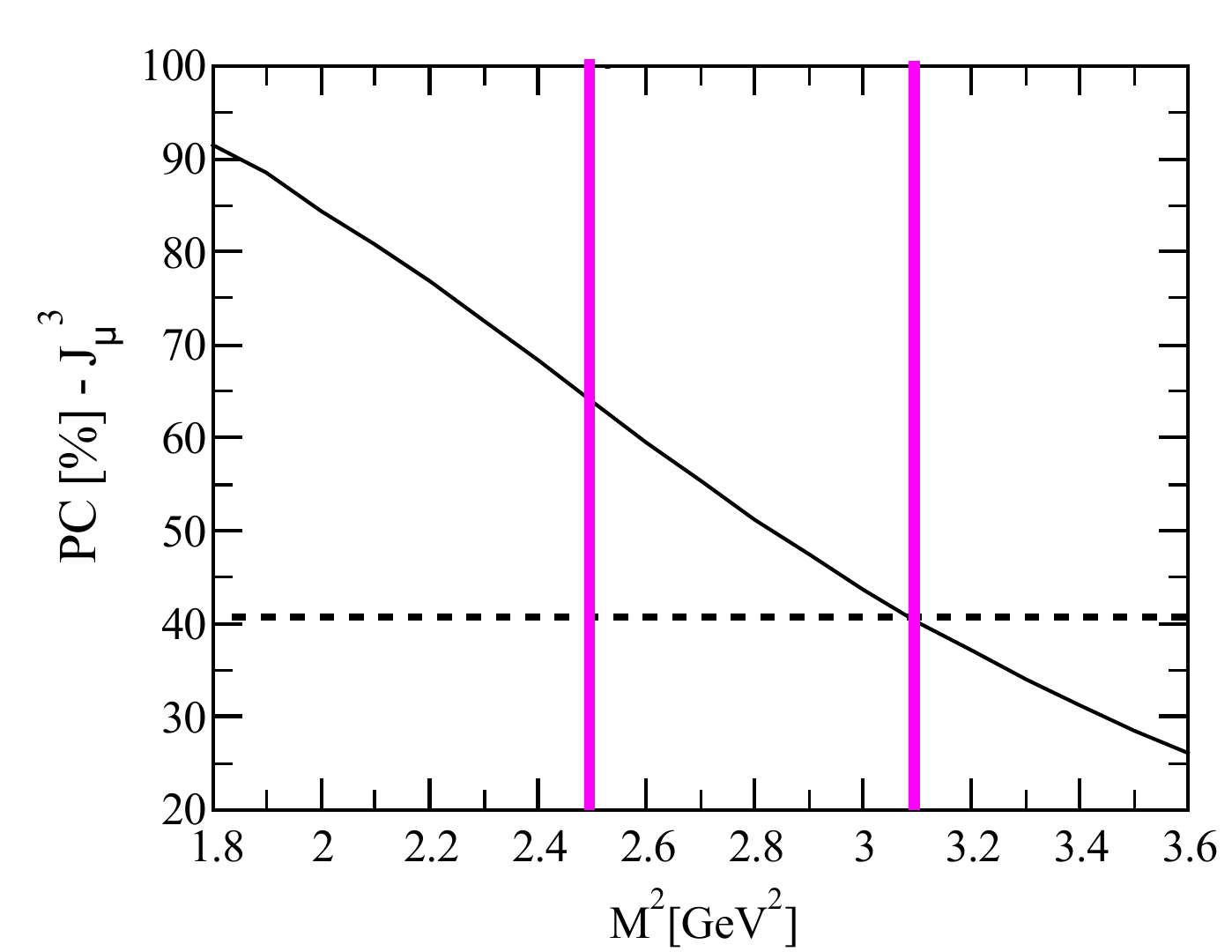}~~ ~~
\includegraphics[width=0.3\textwidth]{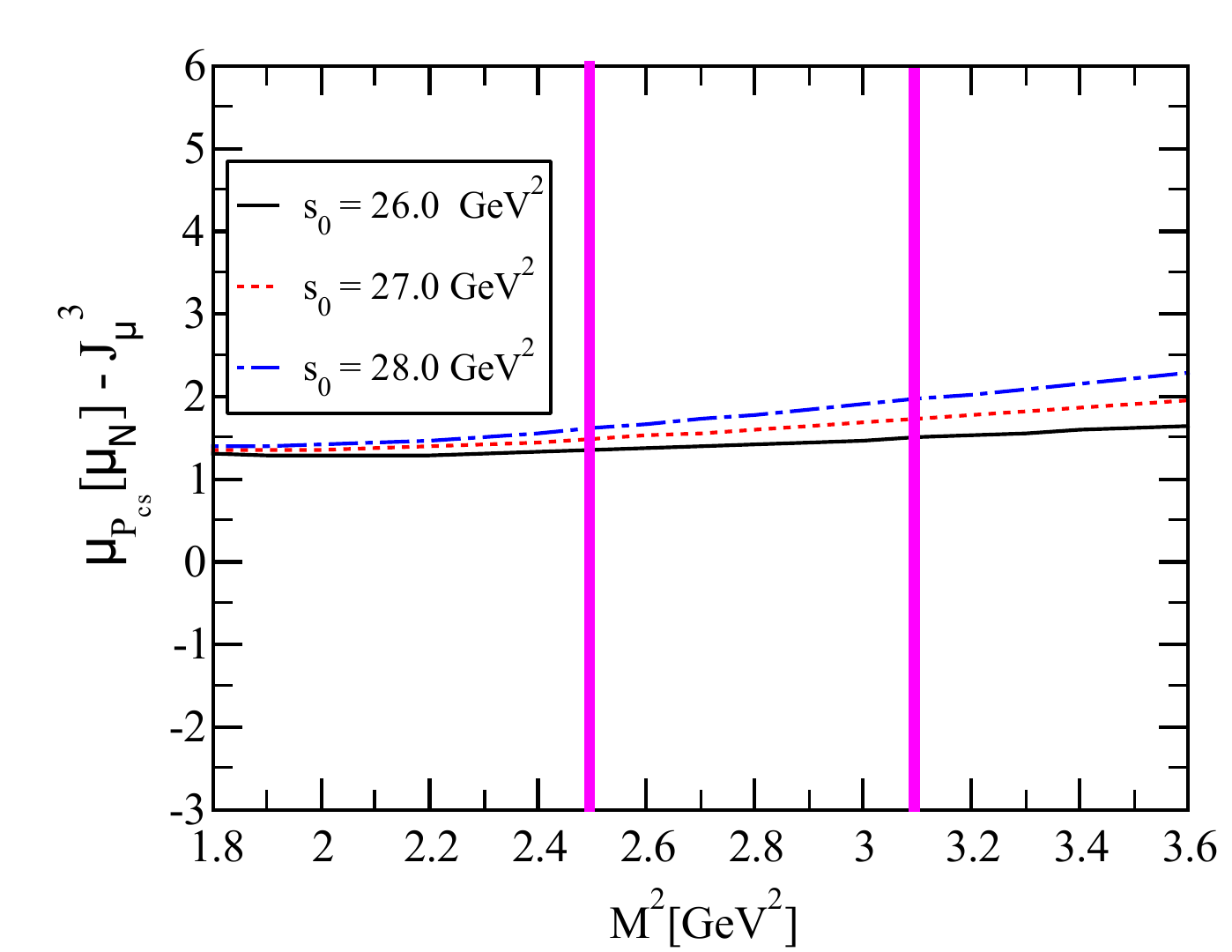}\\

\includegraphics[width=0.3\textwidth]{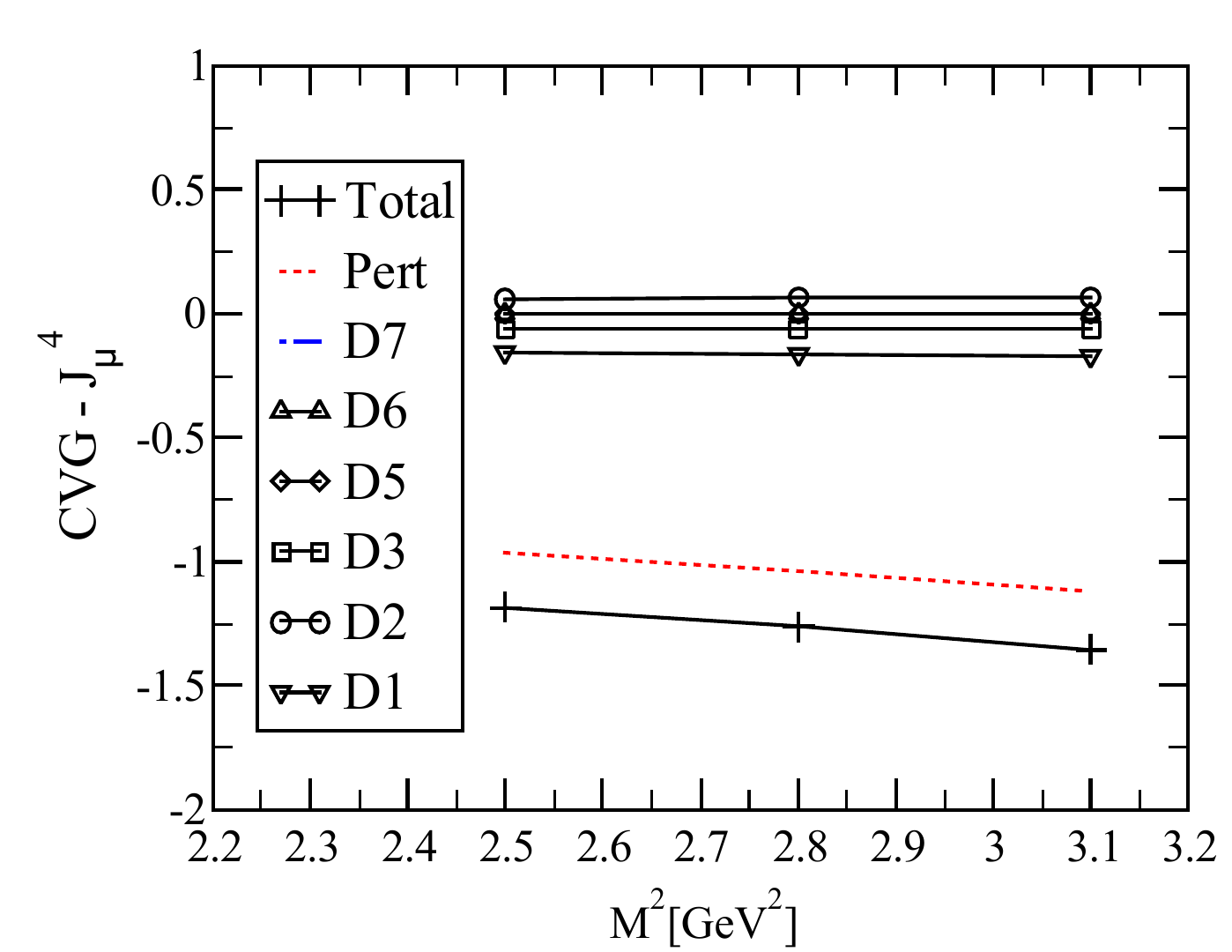}~~ ~~
\includegraphics[width=0.3\textwidth]{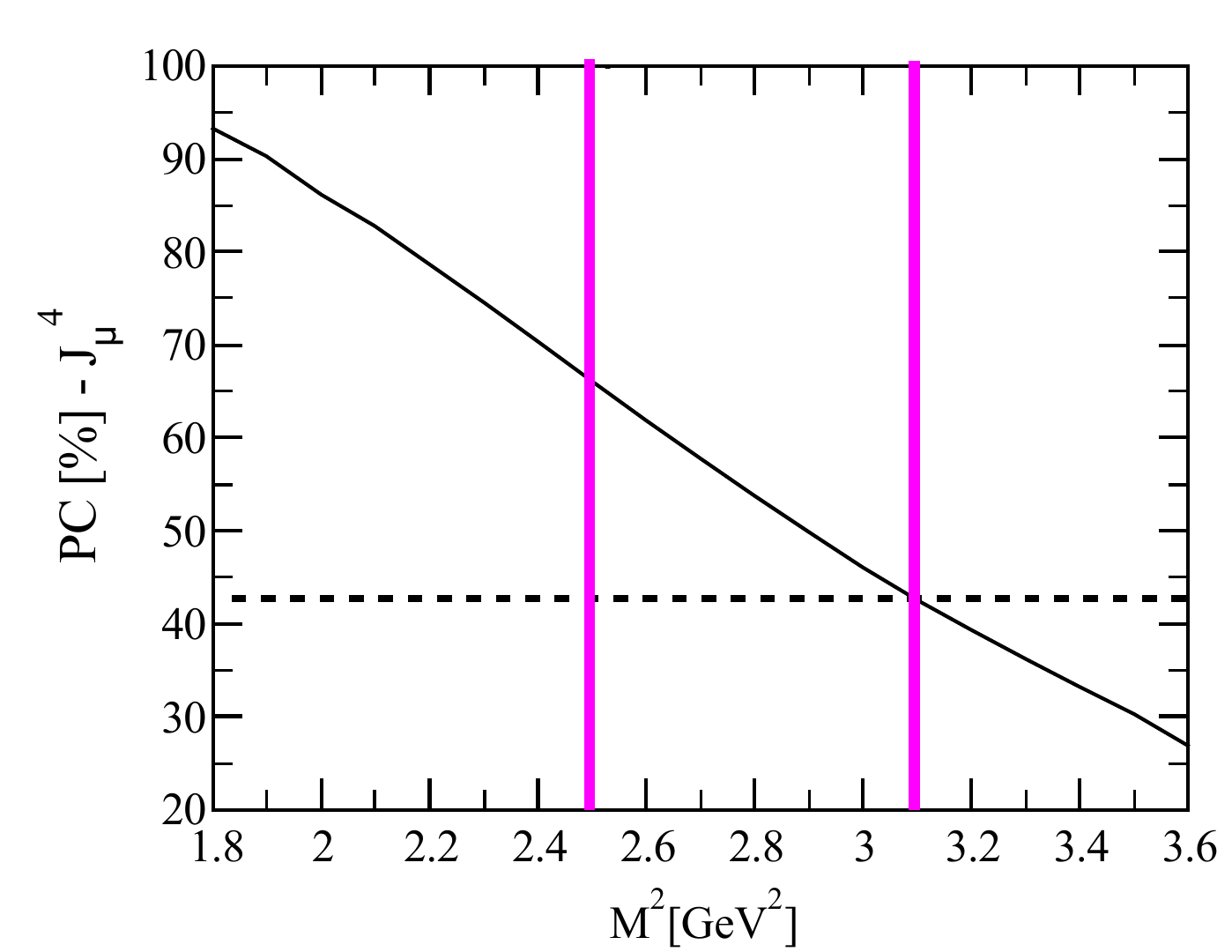}~~ ~~
\includegraphics[width=0.3\textwidth]{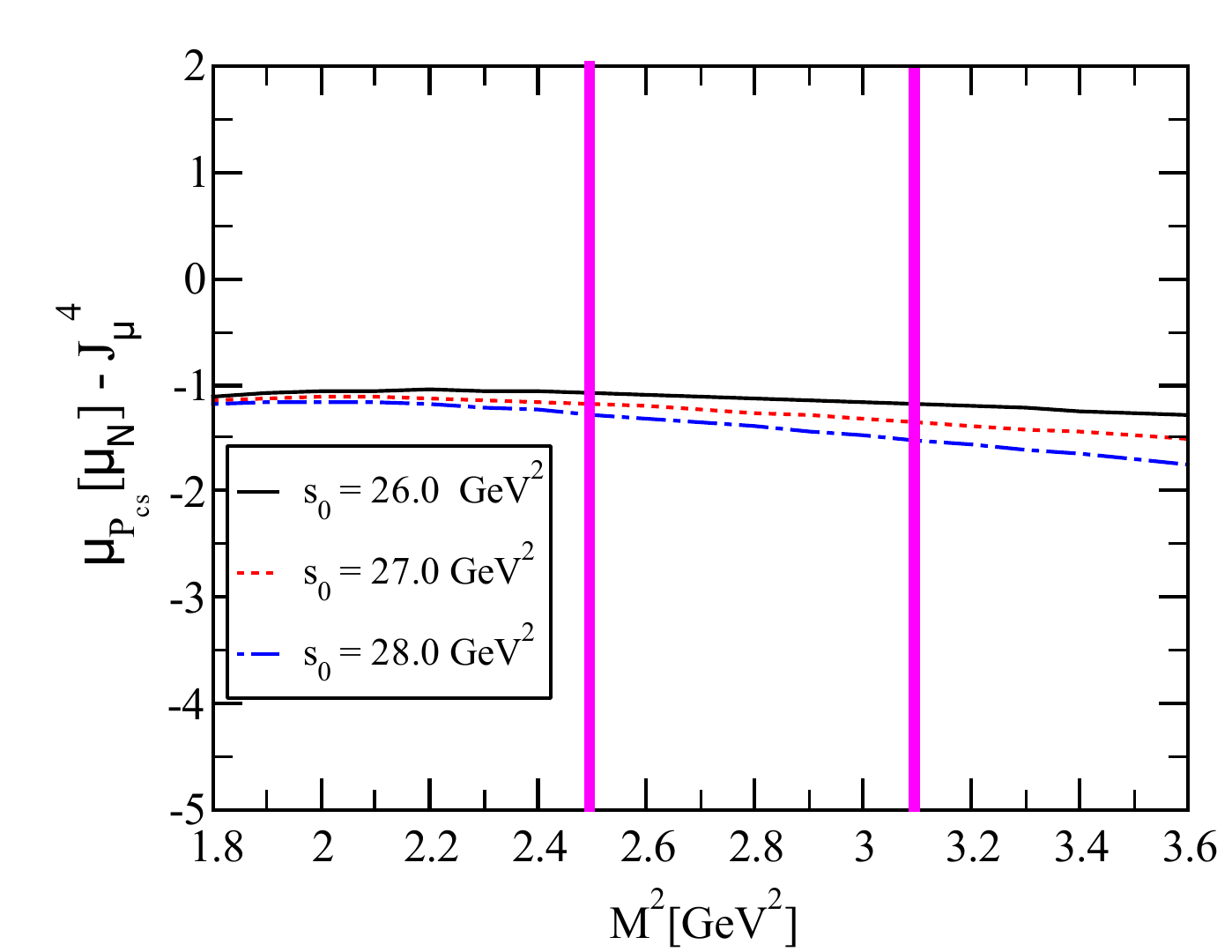}\\

\includegraphics[width=0.3\textwidth]{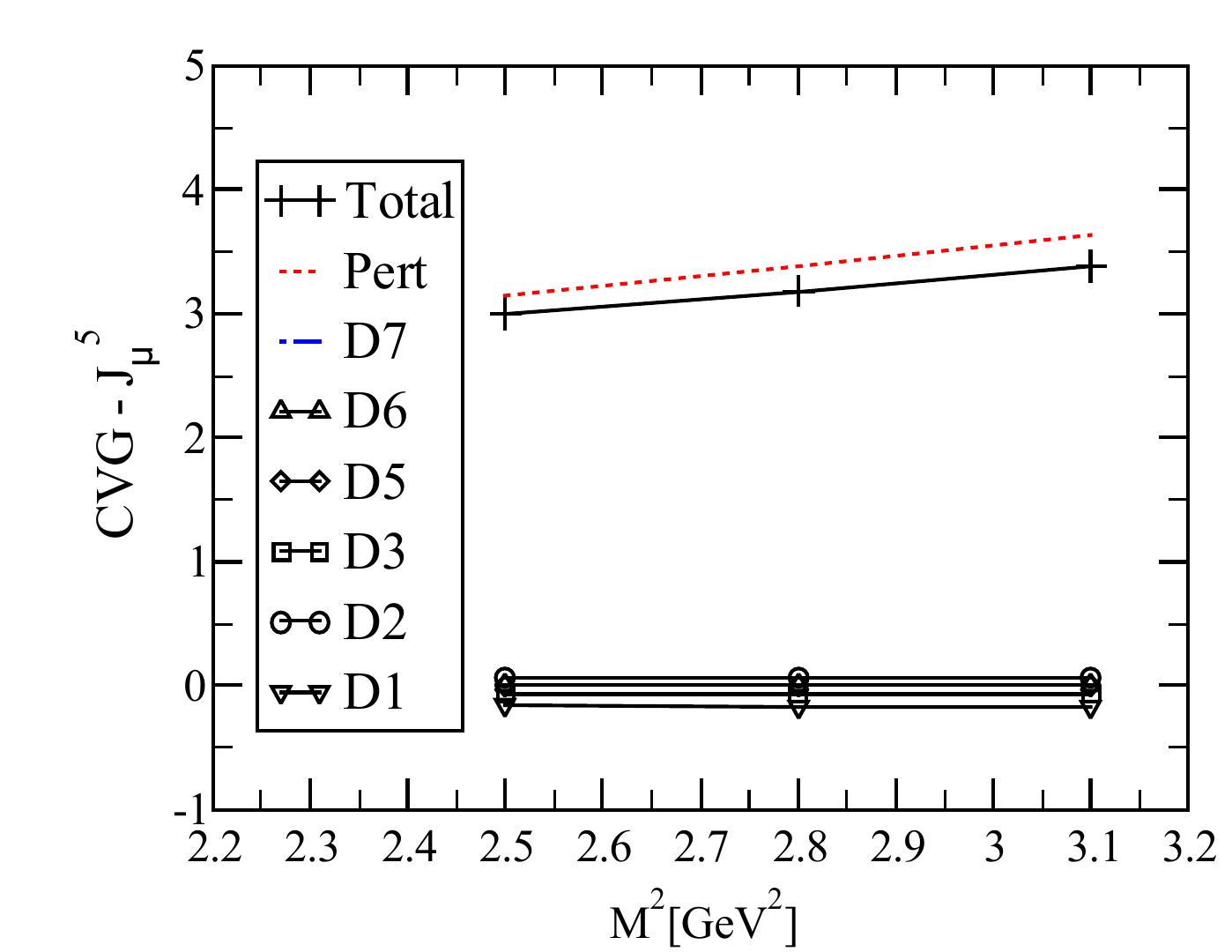}~~ ~~
\includegraphics[width=0.3\textwidth]{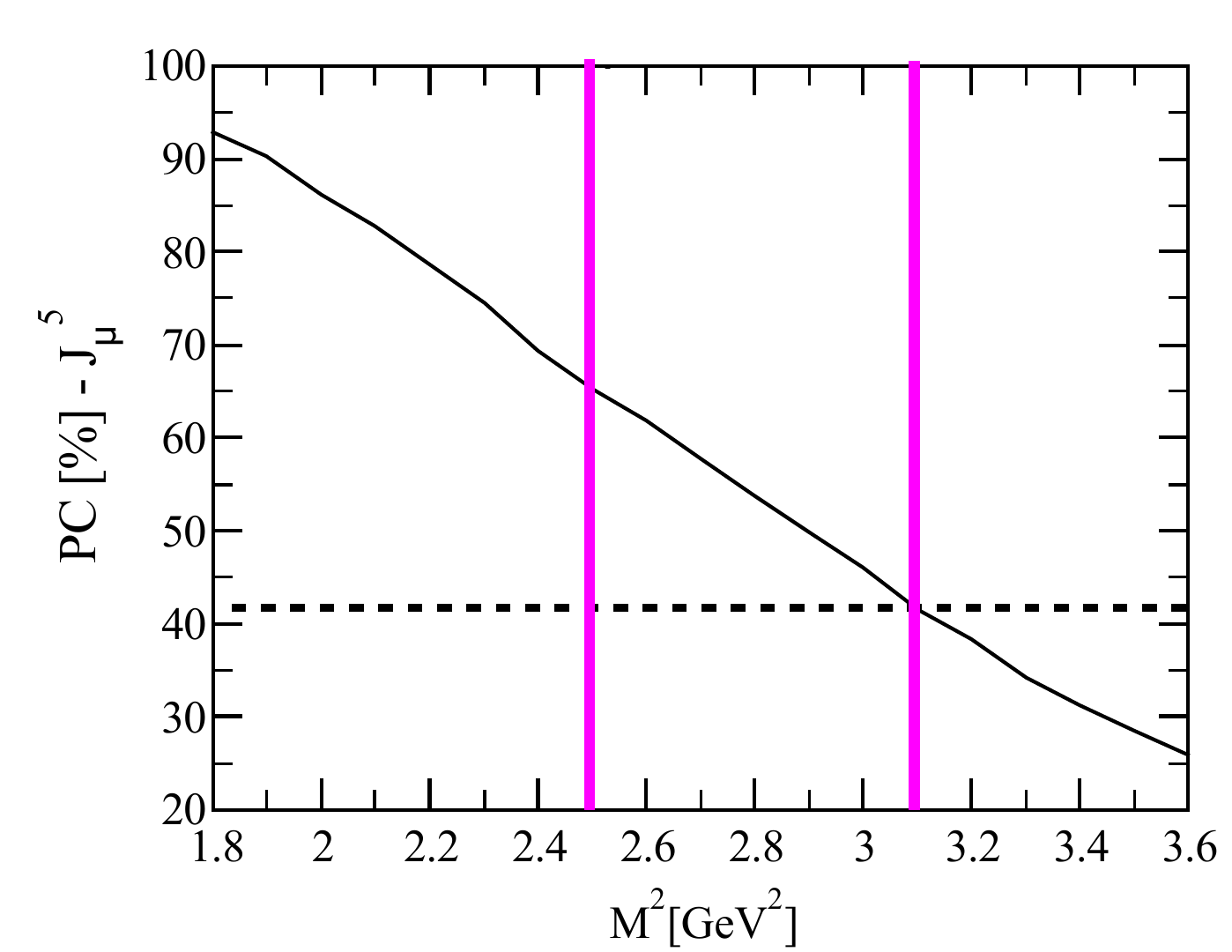}~~ ~~
\includegraphics[width=0.3 \textwidth]{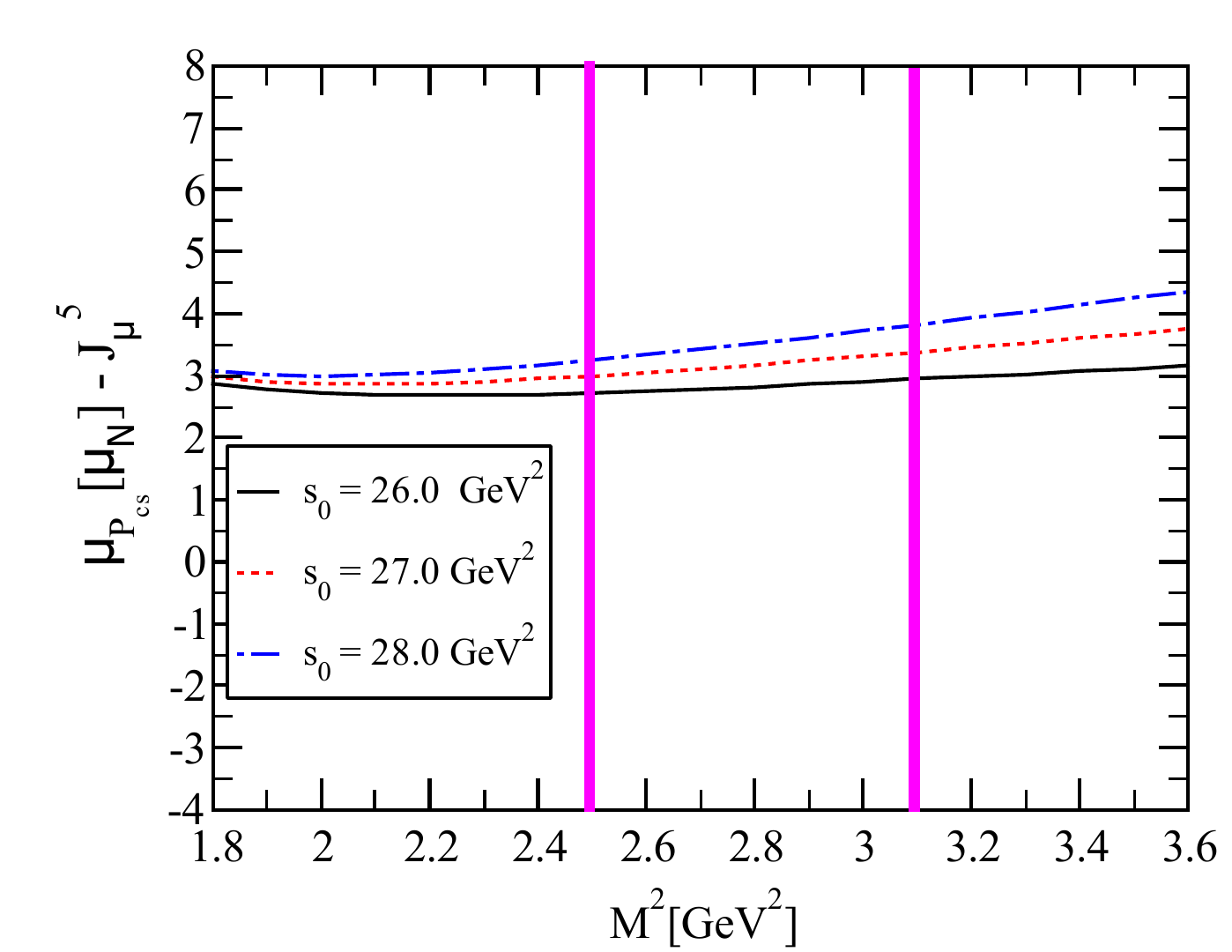}\\
 \caption{CVG analysis (left panels), PC (middle panels), and total value (right panels) for the magnetic dipole moments of the $P^{\Lambda}_{\psi s}$ pentaquarks versus $\rm{M^2}$ at fixed $\rm{s_0}$ values. The adopted Borel window is illustrated by the vertical lines in the middle and right panels, whereas the horizontal line in the middle panel marks the minimum PC value obtained within this region.}
 \label{Msqfig}
  \end{figure}

\begin{figure}[htp]
\centering
\includegraphics[width=0.65\textwidth]{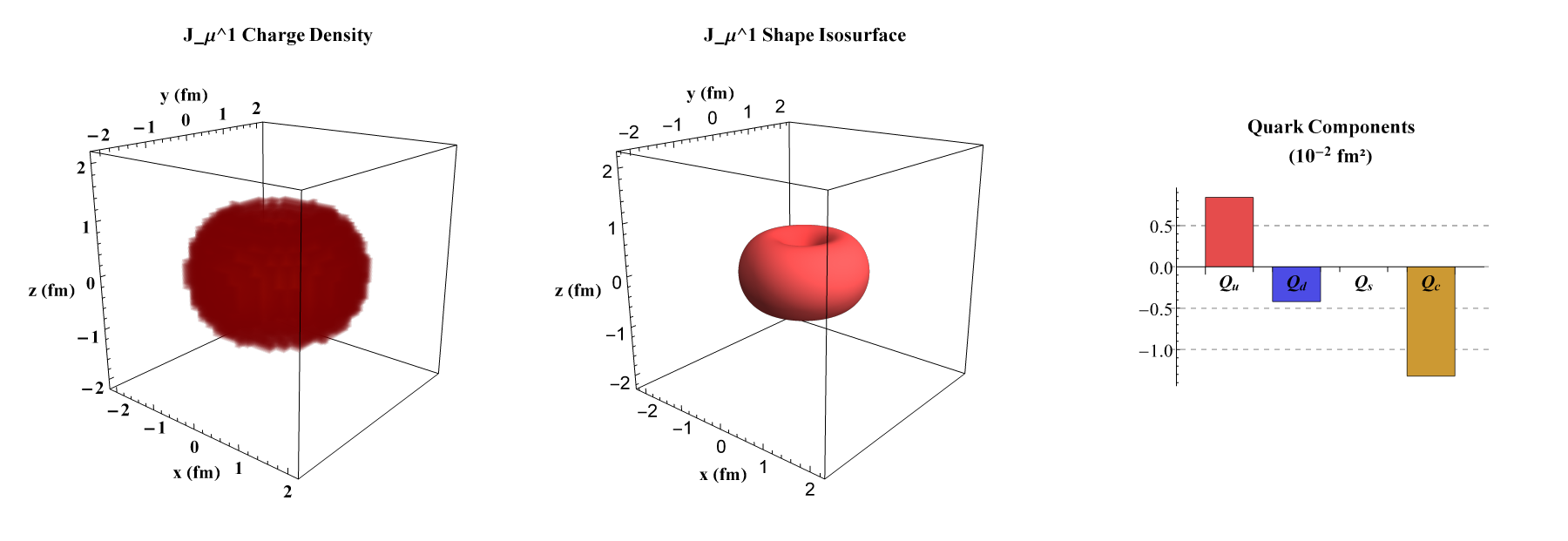}\\
\includegraphics[width=0.65\textwidth]{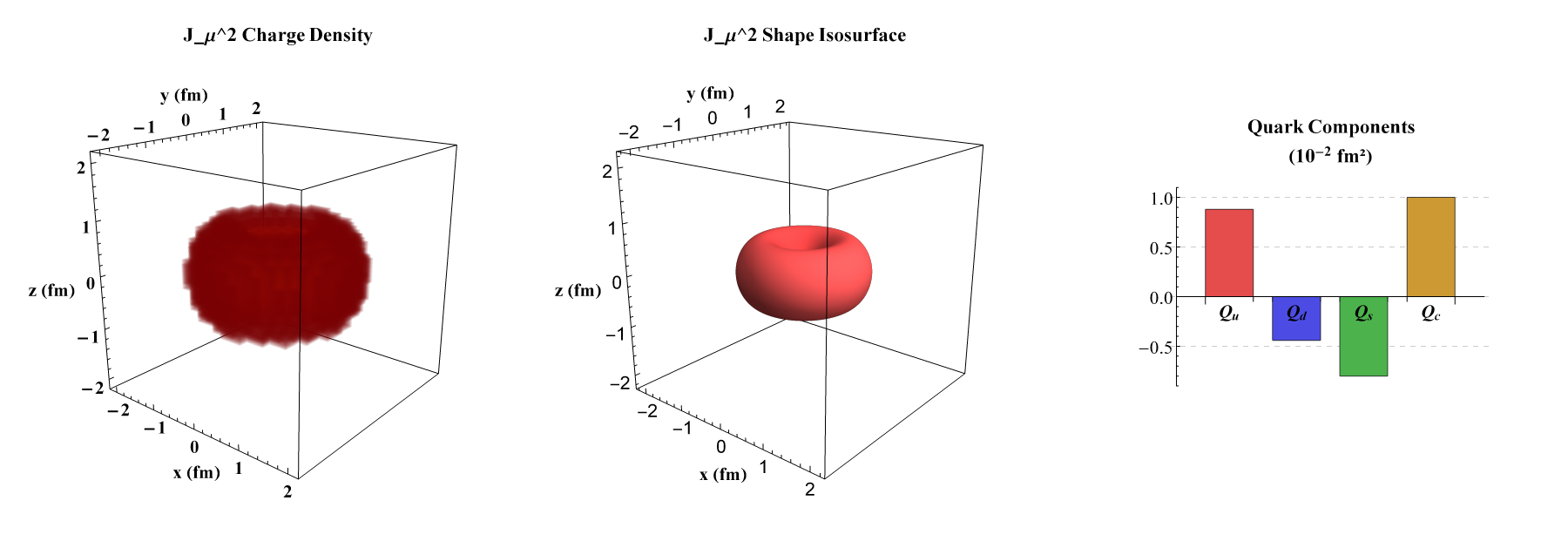}\\
\includegraphics[width=0.65\textwidth]{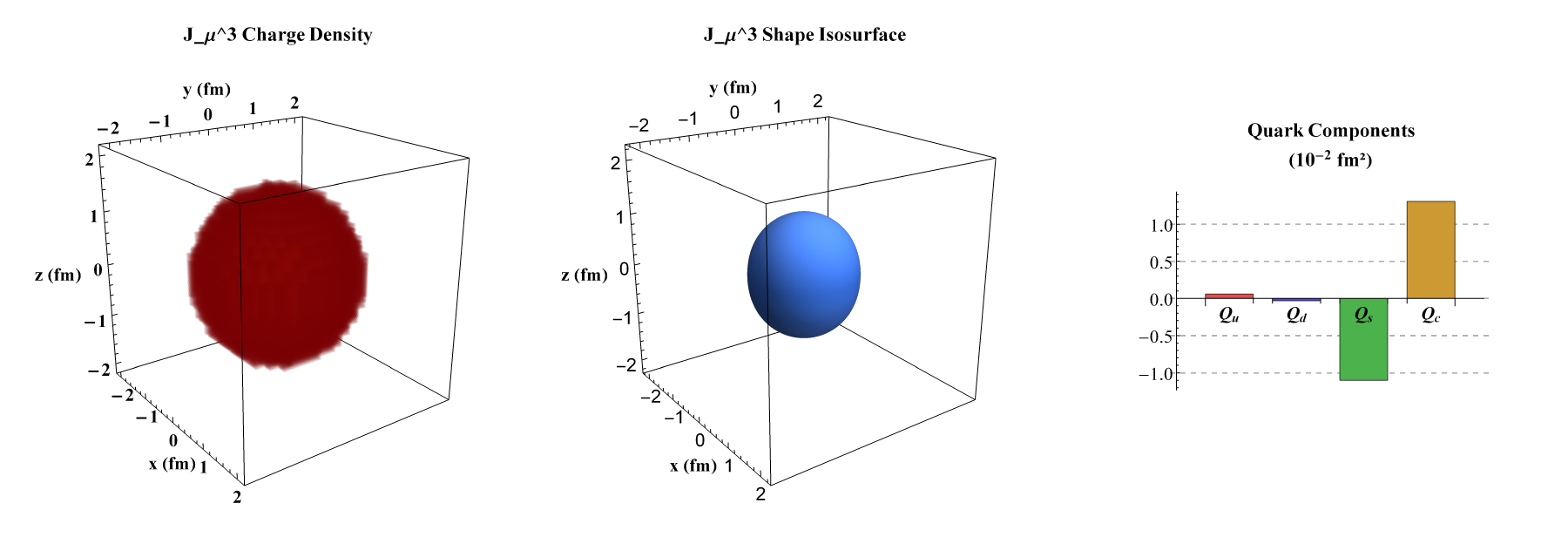}\\
\includegraphics[width=0.65\textwidth]{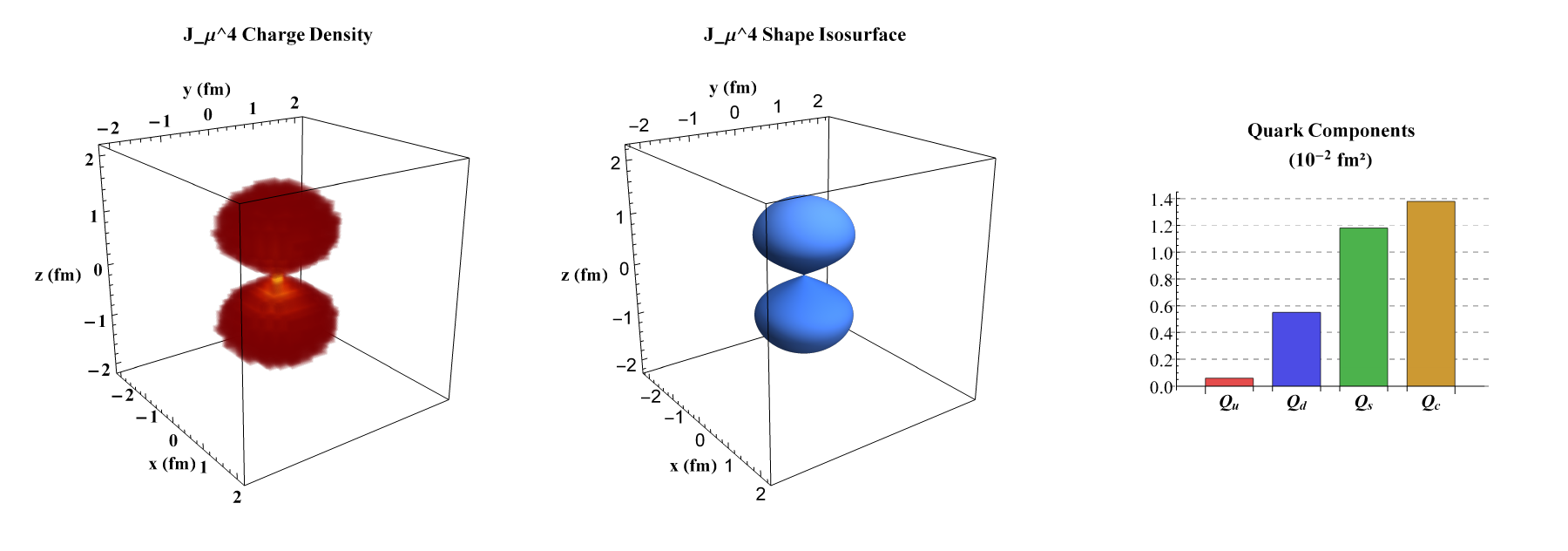}\\
\includegraphics[width=0.65\textwidth]{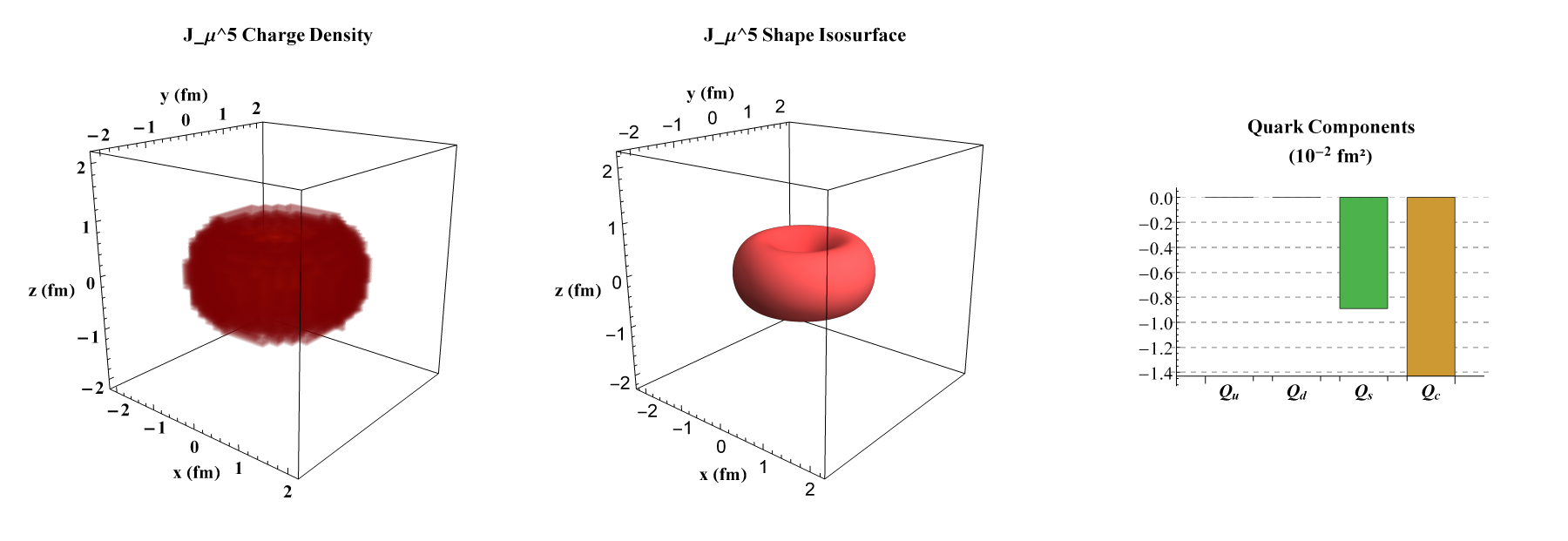}\\
\caption{Analysis of the electric quadrupole  moment for the $P^{\Lambda}_{\psi s}$ pentaquark states.  
Left: three-dimensional charge density distribution $\rho$ (e/fm$^3$); 
Middle: isosurface visualization at 15\% of maximum density showing deformation patterns; 
Right: individual quark contributions to the quadrupole moment (fm$^2$). 
The plots reveal distinct prolate/oblate deformations characteristic of each pentaquark configuration, with color coding indicating density intensity. All spatial axes are in femtometers (fm).}
\label{quark_deformation11}
\end{figure}

\begin{figure}[htp]
\centering
\includegraphics[width=0.85\textwidth]{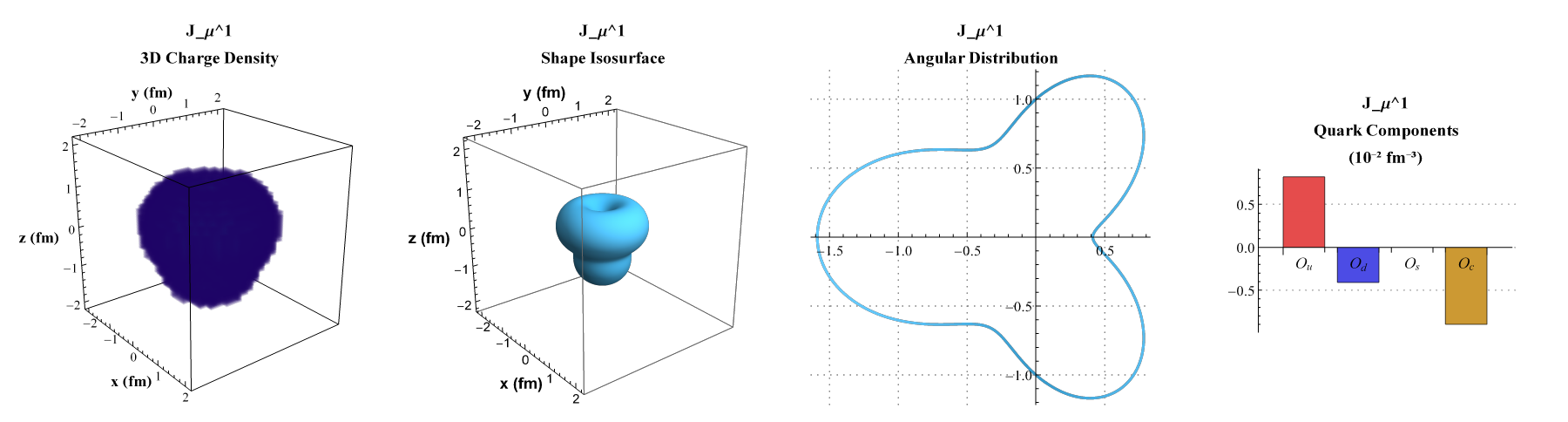}\\
\includegraphics[width=0.85\textwidth]{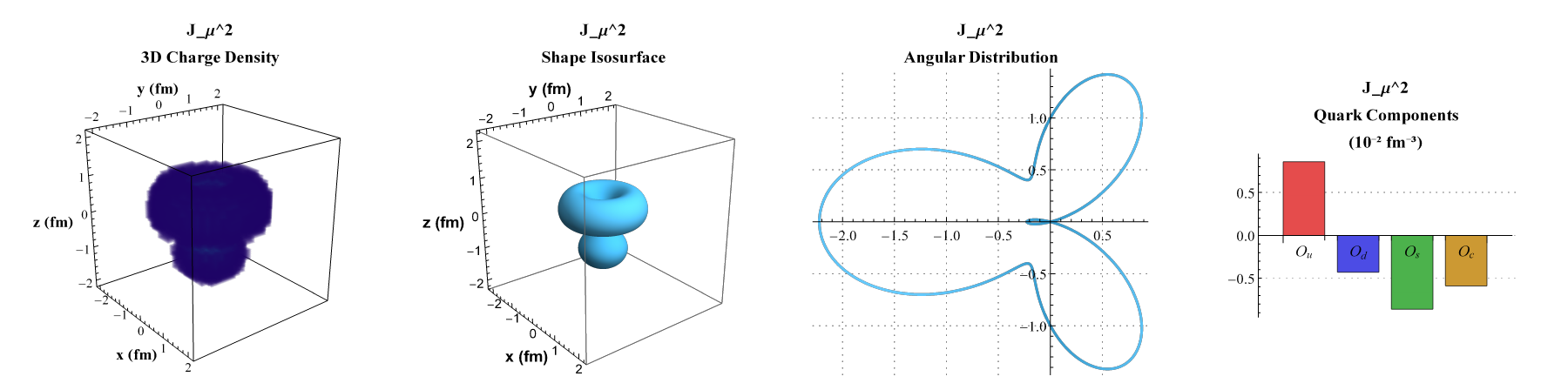}\\
\includegraphics[width=0.85\textwidth]{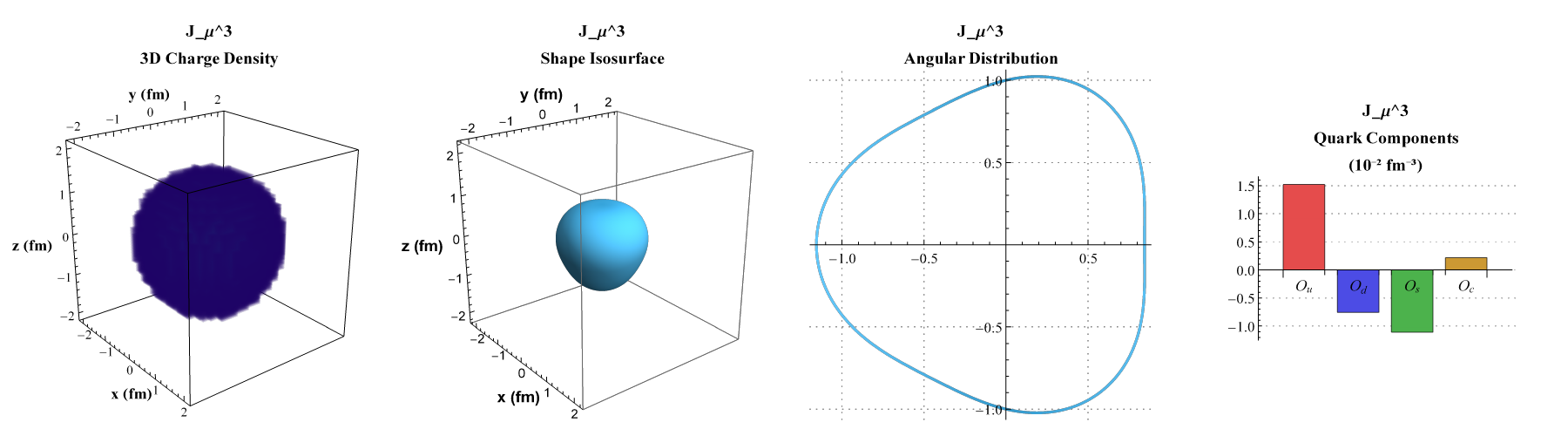}\\
\includegraphics[width=0.85\textwidth]{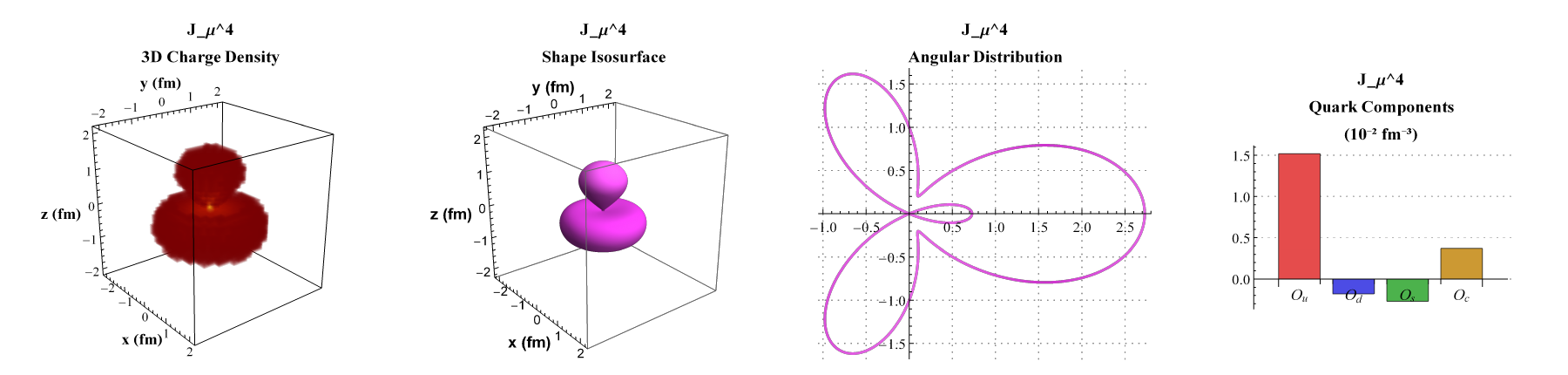}\\
\includegraphics[width=0.85\textwidth]{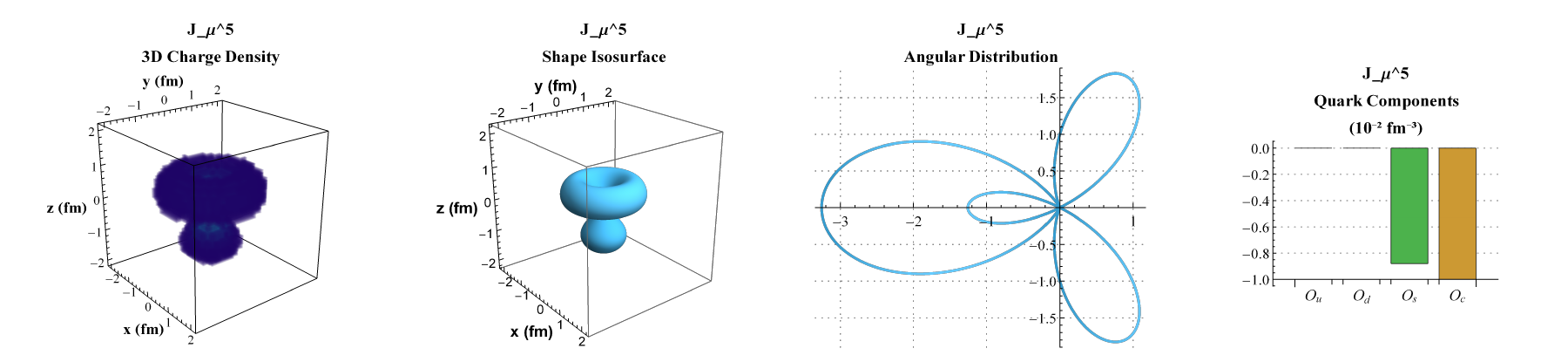}\\
\caption{Analysis of the magnetic octupole moment for the $P^{\Lambda}_{\psi s}$ pentaquark states. 
Left: three-dimensional charge density distribution $\rho$ (e/fm$^3$) exhibiting octupole deformation patterns; 
Middle left: isosurface visualization at 15\% of maximum density, clearly displaying the characteristic peanut shape (positive octupole) or butterfly shape (negative octupole); 
Middle right: angular distribution of charge density showing the $Y_{30} \propto (5\cos^3\theta - 3\cos\theta)$ spherical harmonic dependence; 
Right: quark-sector decomposition of octupole moment components ($\times 10^{-2}$ fm$^3$). 
Positive octupole moments indicate axial charge accumulation (peanut shape), while negative moments show equatorial charge concentration with polar depletion (butterfly shape). The angular distributions reveal the underlying octupole symmetry. All spatial axes are in femtometers (fm).}
\label{quark_deformation12}
\end{figure}

   \end{widetext}

  \newpage

  \begin{widetext} 

 \section*{Appendix: Explicit form of the analytical results for the electromagnetic multipole moments of the $J_\mu^1$ current}
In this appendix, we present the explicit analytical expressions for the magnetic dipole, electric quadrupole, and magnetic octupole moments derived for the $J_\mu^1$ interpolating current. These expressions are obtained after performing the full QCD-side calculation and matching it to the corresponding hadronic representation within the framework of light-cone QCD sum rules. Each multipole moment is expressed in terms of the relevant invariant amplitudes, Borel parameters, and distribution amplitudes, which encode the internal electromagnetic structure of the considered state. The final analytical results are given as follows:
  \begin{align}
   \mu_{ J^1_{\mu}}\, \lambda^2_{J^1_{\mu}} \,e^{-\frac{m^2_{J_\mu^1}}{\mathrm{M^2}}} & =  \rho_1 (\mathrm{M^2},\mathrm{s_0}),\\
   %%%%%%%%%%%%%%%%%%%%%%%%%%%%%%%%%%%%%%%
   \mathcal{Q}_{ J^1_{\mu}}\, \lambda^2_{J^1_{\mu}} \,e^{-\frac{m^2_{J_\mu^1}}{\mathrm{M^2}}}& =  \Delta_1 (\mathrm{M^2},\mathrm{s_0}),\\
     %%%%%%%%%%%%%%%%%%%%%%%%%%%%%%%%%%%%%%%
   \mathcal{O}_{ J^1_{\mu}}\, \lambda^2_{J^1_{\mu}}\, e^{-\frac{m^2_{J_\mu^1}}{\mathrm{M^2}}}& =  \Omega_1 (\mathrm{M^2},\mathrm{s_0}),
  \end{align}
 where, 
\begin{align}
\rho_1 (\mathrm{M^2},\mathrm{s_0})&=\frac{1}{2}\Big[ F_1^{J^1_{\mu}}(\mathrm{M^2},\mathrm{s_0}) +F_2^{J^1_{\mu}}(\mathrm{M^2},\mathrm{s_0})\Big], \\
%%%%%%%%%%%%%%%%%%%%%%%%%%%%%
\Delta_1 (\mathrm{M^2},\mathrm{s_0})&=\Big[ \frac{1}{2}F_1^{J^1_{\mu}}(\mathrm{M^2},\mathrm{s_0}) + m_{ J^1_{\mu}}F_3^{J^1_{\mu}}(\mathrm{M^2},\mathrm{s_0})\Big], \\
       %%%%%%%%%%%%%%%%%%%%%
\Omega_1 (\mathrm{M^2},\mathrm{s_0})&=\Big[ \frac{1}{2}F_1^{J^1_{\mu}}(\mathrm{M^2},\mathrm{s_0}) +\frac{1}{2}F_2^{J^1_{\mu}}(\mathrm{M^2},\mathrm{s_0})+ m_{ J^1_{\mu}}F_3^{J^1_{\mu}}(\mathrm{M^2},\mathrm{s_0})+ 2  m^2_{ J^1_{\mu}}F_4^{J^1_{\mu}}(\mathrm{M^2},\mathrm{s_0})\Big],
      \end{align}
      with
 \begin{align}
F_1^{J^1_{\mu}} (\mathrm{M^2},\mathrm{s_0})&= \frac{1}{2^{28}\times 3 \times 5^3 \times 7 ^2 \pi^7} \Big[(218 e_c -621 (e_d + e_u)) I[0, 7] +   (378 e_c -1113 (e_d + e_u))  I[1, 6]  
 \Big]
  %%%%%%%%%%%%%%%%%%%%%%%%%%%%%%%%%%%%%%%%%%%%%%%%%%%%%%%%%%%
 \nonumber\\
 &+ \frac{ \langle g_s^2 G^2\rangle \langle \bar q q \rangle }{2^{28}\times 3^6 \times 5  \pi^5}  \Big[  \Big (-160 (e_d + e_u) m_c \mathbb A[u_ 0] - 
    99 e_u m_s I_ 1[\mathcal S] - 342 e_u m_s I_3[\mathcal S] +  
    828 ((e_d + e_u) m_c \nonumber\\
    &- e_d m_s) \big (3 I_ 3[\mathcal  T_ 1] + 
        2 (I_ 3[\mathcal T_ 2] + I_ 3[\mathcal T_ 4])\big)\Big) I[0, 
   3] + 64 \chi (e_d + e_u) m_c I[0, 4] \varphi_\gamma[u_ 0]  \Big]
   \nonumber\\
  %%%%%%%%%%%%%%%%%%%%%%%%%%%%%%%%%%%%%%%%%%%%%%%%%%%%%%%%%%%
 % \end{align}
 % \begin{align}
 &+ \frac{11 \langle g_s^2 G^2\rangle f_{3\gamma}}{2^{31}\times 3^6 \times 5  \pi^5}
 \Big[  891 (2 e_s + e_u) I_ 1[\mathcal A] + 
 792 (2 e_s + e_u) I_ 1[\mathcal V] + 1782 e_s I_ 3[\mathcal A] + 
 2961 e_u I_ 3[\mathcal A] - 1584 e_s I_ 3[\mathcal V] \nonumber\\
    &- 
 2862 e_u I_ 3[\mathcal V] - 144 e_u I_ 5[\psi^a] - 
 504 e_s \psi^a[u_ 0] - 152 e_u \psi^a[u_ 0] + 
 104 e_u \psi_\gamma^\nu[u_ 0] + 
 2 e_d (1863 I_ 3[\mathcal A] - 1863 I_ 3[\mathcal V] \nonumber\\
    &  - 
    252 I_ 5[\psi^a]+ 5 \psi^a[u_ 0] + 178 \psi_\gamma^\nu[u_ 0])
 \Big]
  \nonumber\\
  %%%%%%%%%%%%%%%%%%%%%%%%%%%%%%%%%%%%%%%%%%%%%%%%%%%%%%%%%%%
 &+ \frac{\langle \bar q q \rangle}{2^{25}\times 3 \times 5^2  \pi^5}\Big[ \Big(-5 e_u m_s I_3[\mathcal S] + 
   2 (2 (e_d + e_u) m_c - 5 e_d m_s) (3 I_3[\mathcal T_1] + 
      2 (I_3[\mathcal T_2] + I_3[\mathcal T_4]))\Big) I[0, 5]\Big]
\nonumber\\
  %%%%%%%%%%%%%%%%%%%%%%%%%%%%%%%%%%%%%%%%%%%%%%%%%%%%%%%%%%%
 &+ \frac{f_{3\gamma}}{2^{26}\times 3^2 \times 5  \pi^5}\Big[ 
(2 e_d + e_u) (I_3[\mathcal A] - I_3[\mathcal V]) I[0, 6]\Big],\label{F1sonuc}%\\
%\nonumber\\
 \end{align}
\begin{align}
F_2^{J^1_{\mu}} (\mathrm{M^2},\mathrm{s_0})&=- \frac{9 e_c}{2^{26} \times 5^3 \times 7  \pi^7} I[1, 6] 
  %%%%%%%%%%%%%%%%%%%%%%%%%%%%%%%%%%%%%%%%%%%%%%%%%%%%%%%%%%%
 \nonumber\\
 &+ \frac{ \langle g_s^2 G^2\rangle \langle \bar q q \rangle }{2^{26}\times 3^6 \times 5  \pi^5}  \Big[ \Big (80 (e_d + e_u) m_c \mathbb A[u_0] + 
    9 e_u m_s (13 I_ 1[\mathcal S] - 11 I_ 3[\mathcal S])\Big) I[0, 
   3] - 32 \chi (e_d + e_u) \nonumber\\
   &\times m_c I[0, 4] \varphi_\gamma[u_ 0]  \Big]
   \nonumber\\
  %%%%%%%%%%%%%%%%%%%%%%%%%%%%%%%%%%%%%%%%%%%%%%%%%%%%%%%%%%%
 &+ \frac{\langle g_s^2 G^2\rangle f_{3\gamma}}{2^{31}\times 3^5 \times 5  \pi^5}
 \Big[  \Big(  69 (36 e_d - 14 e_s + 13 e_u) I_1[\mathcal V] - 
 12 (207 e_d - 58 e_s + 86 e_u) I_3[\mathcal V] + 
 48 (14 e_d + 10 e_s \nonumber\\
 &+ 9 e_u) I_5[\psi^a] - 4 (61 e_d + 34 e_u) \psi^a[u_0] \Big)I[0,4] \Big]
  \nonumber\\
  %%%%%%%%%%%%%%%%%%%%%%%%%%%%%%%%%%%%%%%%%%%%%%%%%%%%%%%%%%%
 &+ \frac{e_u\,m_s\,\langle \bar q q \rangle}{2^{26}\times 3 \times 5  \pi^5}  I_1[\mathcal S]\,   I[0, 5]
\nonumber\\
  %%%%%%%%%%%%%%%%%%%%%%%%%%%%%%%%%%%%%%%%%%%%%%%%%%%%%%%%%%%
 &+ \frac{f_{3\gamma}}{2^{28}\times 3^3 \times 5  \pi^5}\Big[\Big((48 e_d + 2 e_s + 25 e_u) I_1[\mathcal V] - 4 (12 e_d + 2 e_s + 7 e_u) I_3[\mathcal V]\Big) I[0, 6]\Big],\label{F2sonuc}%\\
 %\nonumber\\
 \end{align}
 \begin{align}
F_3^{J^1_{\mu}} (\mathrm{M^2},\mathrm{s_0})&= \frac{191 \, e_c\,m_c}{2^{25} \times 3^2 \times 5^3 \times 7  \pi^7} I[0, 6] 
  %%%%%%%%%%%%%%%%%%%%%%%%%%%%%%%%%%%%%%%%%%%%%%%%%%%%%%%%%%%
 \nonumber\\
 &+ \frac{ \langle g_s^2 G^2\rangle \langle \bar q q \rangle }{2^{26}\times 3^6 \times 5  \pi^5}  \Big[ e_u m_c m_s I_3[\mathcal S] I[0, 2]  \Big]
   \nonumber\\
  %%%%%%%%%%%%%%%%%%%%%%%%%%%%%%%%%%%%%%%%%%%%%%%%%%%%%%%%%%%
 &+ \frac{\langle g_s^2 G^2\rangle f_{3\gamma}\,m_c}{2^{29}\times 3^4 \times 5  \pi^5}
 \Big[  \Big( (187 e_d - 40 e_s + 112 e_u) I_1[\mathcal V] - 30 (55 e_d + 28 e_u) I_3[\mathcal V] - 
 2 (2 e_s + e_u) (36 I_5[\psi^a] \nonumber\\
 &+ 53 \psi^a[u_0]) \Big)I[0,3] \Big]
  \nonumber\\
  %%%%%%%%%%%%%%%%%%%%%%%%%%%%%%%%%%%%%%%%%%%%%%%%%%%%%%%%%%%
 &+ \frac{e_u\,m_s\,m_c\, \langle \bar q q \rangle}{2^{23} \times 5  \pi^5}  I_3[\mathcal S]\,   I[0, 4]
\nonumber\\
  %%%%%%%%%%%%%%%%%%%%%%%%%%%%%%%%%%%%%%%%%%%%%%%%%%%%%%%%%%%
 &- \frac{f_{3\gamma}\, m_c }{2^{24}\times 3 \times 5  \pi^5}\Big[((2 e_s + e_u) I_1[\mathcal V] + 1_2 (2 e_d + e_u) I_3[\mathcal V]) I[0, 5]\Big],\label{F3sonuc}  \\
 \nonumber\\ 
 %%%%%%%%%%%%%%%%%%%%%%%%%%%%%%%%%%%%%%%%%%%%%%%%%%%%%%%%%%%%%%%%%%%%%%%%%%%%%%%%%%%%%%%%%%%%%%%%%%%%%%%%%%%%%%%%%%%%%%%%%%%%%%%
 F_4^{J^1_{\mu}} (\mathrm{M^2},\mathrm{s_0})&= \frac{\langle g_s^2 G^2\rangle f_{3\gamma}\,m_c}{2^{27}\times 3^3 \times 5  \pi^5}
 \Big[    (11 (17 e_d + 12 e_u) I_3[\mathcal V] + 7 (2 e_s + e_u) \psi^a[u_0]) I[0, 2]\Big],
 \label{F4sonuc} 
 \end{align}
 \noindent  where the functions $\mathrm{I}[n,m]$, $ I_1[\mathcal{F}]$, $ I_3[\mathcal{F}]$, and $ I_5[\mathcal{F}]$  are listed as      
\begin{align}
 \mathrm{I}[n,m]&= \int_{\mathcal M}^{\mathrm{s_0}} ds~ e^{-s/\mathrm{M^2}}~
 s^n\,(s-\mathcal M)^m,\nonumber\\
 I_1[\mathcal{F}]&=\int D_{\alpha_i} \int_0^1 dv~ \mathcal{F}(\alpha_{\bar q},\alpha_q,\alpha_g)  \delta'(\alpha_ q +\bar v \alpha_g-u_0),\nonumber\\
     I_3[\mathcal{F}]&=\int D_{\alpha_i} \int_0^1 dv~ \mathcal{F}(\alpha_{\bar q},\alpha_q,\alpha_g)
 \delta(\alpha_ q +\bar v \alpha_g-u_0),\nonumber\\
    I_5[\mathcal{F}]&=\int_0^1 du~ \mathcal{F}(u)\delta'(u-u_0),
 \end{align}
\noindent where $\mathcal{M} = (2m_c + m_s)^2$, and $\mathcal{F}$ denotes the photon DAs.

The Borel transformation relations employed in the present analysis can be expressed as
\begin{equation}
\mathcal{B}\!\left\{\frac{1}{\big[[p^2 - m_i^2][(p + q)^2 - m_f^2]\big]}\right\} 
\!\rightarrow\! e^{-m_i^2/M_1^2 - m_f^2/M_2^2},
\end{equation}
on the hadronic representation, and
\begin{equation}
\mathcal{B}\!\left\{\frac{1}{\big(m^2 - \bar{u}p^2 - u(p + q)^2\big)^{\alpha}}\right\} 
\!\rightarrow\! (M^2)^{2 - \alpha}\, \delta(u - u_0)\, e^{-m^2/M^2},
\end{equation}
on the QCD side of the correlation function. The auxiliary parameters entering these transformations are defined as
\begin{equation*}
M^2 = \frac{M_1^2 M_2^2}{M_1^2 + M_2^2}, \qquad
u_0 = \frac{M_1^2}{M_1^2 + M_2^2}.
\end{equation*}
Here, $M_1^2$ and $M_2^2$ denote the Borel mass parameters associated with the initial and final $P^{\Lambda}_{\psi s}$ states, respectively. 
These parameters arise as a consequence of applying independent Borel transformations to the initial and final hadronic momenta, $p^2$ and $(p+q)^2$, in order to exponentially suppress contributions from higher-mass states and the continuum part of the spectral density. 
Since the same physical state $P^{\Lambda}_{\psi s}$ contributes at both the initial and final vertices of the correlation function, the most natural and physically consistent choice is to set $M_1^2 = M_2^2 = 2M^2$, which immediately yields $u_0 = \tfrac{1}{2}$. This symmetric choice ensures that the single Borel parameter $M^2$ effectively represents the average virtuality of the heavy--light system and optimally stabilizes the QCD sum rule against numerical uncertainties. Moreover, this parametrization guarantees that the single dispersion integral used in the sum-rule formalism remains convergent and that the ground-state contribution dominates over those from excited resonances and the continuum threshold. In other words, the adopted equality of the Borel parameters provides a balanced exponential suppression on both sides of the correlator, improving the reliability and stability of the extracted form factors and multipole moments. A comprehensive account of this procedure and its applications to heavy exotic systems can be found in Ref.~\cite{Ozdem:2024dbq}.

\end{widetext}

\bibliographystyle{elsarticle-num}
\bibliography{PcsMM.bib}
\end{document}